\documentclass[aps, prd, showkeys, showpacs]{revtex4}

\usepackage{amsmath,amsfonts,amscd,amssymb,epsf}
\usepackage[small]{caption}

\newcommand{\pa}{\partial}

\begin{document}

\title{Casimir effect  of electromagnetic field in $\boldsymbol{D}$-dimensional spherically symmetric cavities}

 \author{L.P. Teo}\email{ LeePeng.Teo@nottingham.edu.my}\address{Department of Applied Mathematics, Faculty of Engineering, University of Nottingham Malaysia Campus, Jalan Broga, 43500, Semenyih, Selangor Darul Ehsan, Malysia. }

\begin{abstract}
 Eigenmodes of electromagnetic field with perfectly conducting or infinitely permeable  conditions on the boundary of a $D$-dimensional spherically symmetric cavity is derived explicitly. It is shown that there are $(D-2)$ polarizations   for TE modes and one polarization  for TM modes, giving rise to a total of $(D-1)$ polarizations.  In case of a $D$-dimensional ball, the eigenfrequencies of electromagnetic field with perfectly conducting boundary condition coincides with the eigenfrequencies of gauge  one-forms with relative boundary condition; whereas the eigenfrequencies of electromagnetic field with infinitely permeable boundary condition coincides with the eigenfrequencies of gauge  one-forms with absolute boundary condition. Casimir energy for a $D$-dimensional spherical shell configuration is computed using both cut-off regularization and zeta regularization. For a double spherical shell configuration, it is shown that the Casimir energy can be written as a sum of the single spherical shell contributions and an interacting term, and the latter is free of divergence. The interacting term always gives rise to an attractive force between the two spherical shells.  Its   leading term is the Casimir force acting between two parallel plates of the same area, as expected by proximity  force approximation.

\end{abstract}
 \maketitle

\section{Introduction}

This decade has observed a  tremendous progress in the research in Casimir effect. One of the motivations comes from the wide applications of Casimir effect in other areas of physics \cite{1}. Besides, successful experimental verifications of this quantum effect by various groups of researchers have greatly stimulated the  research in this area. Before the turn of this century, the research in Casimir effect focused on simple geometric configurations such as parallel plates, rectangular, spherical and cylindrical geometries, where specific coordinate systems are available for explicit solutions of  the corresponding energy spectrums. For more complicated configurations such as sphere-plate configuration, approximation methods like proximity force approximation have to be employed to approximate the Casimir force. In this decade, more attention was paid to   numerically calculate the Casimir energy or Casimir force between arbitrary compact objects. Various methods have been proposed such as the semiclassical method, the worldline method and the multiple scattering approach.

Most of the works in Casimir effect, especially those involve real materials considered  $(3+1)$-dimensional Minskowski spacetime.  Nevertheless,  the exploration of Casimir effect in higher dimensional spacetime has a long history. One of the pioneering works is the work by Ambj\o rn and Wolfram \cite{2}, where they derived Casimir energy for massless scalar field and massless vector field (electromagnetic field) in a $D$-dimensional rectangular cavity. Despite of the long history, the research on Casimir effect in higher dimensional spacetime has been focused on massless scalar field. Relatively few works have considered massless vector field since the eigenmodes are harder to solve compared to the scalar case. Very often the results for vector field were incorrectly claimed as a simple multiple of the results for scalar field. Except for the configuration of a pair of parallel hyperplanes, this claim is not true. In case of $D$-dimensional rectangular cavity, this can be seen from the results of \cite{2}.

Motivated by Casimir's intriguing model for electron \cite{3}, Boyer \cite{4} calculated the Casimir force acting on a perfectly conducting three dimensional spherical shell and found that it is repulsive. This result has later been confirmed by several groups of researchers \cite{5,6,7,35,8,9,36,12} using different  methods.  In this case, the eigenmodes can be divided into transverse electric (TE) modes and transverse magnetic (TM) modes. Except for the absence of the $l=0$ modes, the TE modes is the same as the  eigenmodes for massless scalar field with Dirichlet boundary conditions, and the TM modes correspond to the eigenmodes of massless scalar field with Robin (mixed) boundary conditions with a suitable Robin parameter.  For massive scalar field confined in the interior and exterior of a three dimensional spherical surface, the Casimir energies are computed in \cite{15}.  The extension to   higher dimensions was first carried out in \cite{10},  where the Casimir stress acting on a $D$-dimensional spherical shell was computed for massless scalar field using Green's function method.   This was later reconsidered in \cite{14} using zeta function method. An attempt to generalize this result to massless vector field was carried out in \cite{13,11}. By analogy with the $D=3$ case,    the Casimir stress acting on a $D$-dimensional perfectly conducting spherical shell is divided into contribution from TE modes and   TM modes, where the contribution by TE modes is equated to the contribution by massless scalar field with Dirichlet boundary conditions after the omission of the $l=0$ mode, and the contribution by TM modes is equated to the contribution by massless scalar field with Robin boundary conditions with a suitable Robin parameter. Later in \cite{16}, a systematic approach for  scalar field, spinor field and electromagnetic field in $D$-dimensional spherical cavity was given, and the numerical results for Casimir energies when $2\leq D\leq 9$ were tabulated for these different quantum fields. Although the results for massless scalar field in $D$-dimensional spherical cavity is authentic, there are some reasonable doubts that can be raised on  the results for electromagnetic field. In the work \cite{2}, it was found that the electromagnetic field in a $D$-dimensional cavity has $(D-1)$ polarizations. However in \cite{13,11,16}, the electromagnetic field only has two   polarizations in a $D$-dimensional spherical cavitiy.   In our opinion, this inconsistency is because the electromagnetic field in $D$-dimensional spherical cavity was considered in \cite{13,11,16} as a pure direct generalization of the $D=3$ case, without reference to the equation of motion satisfied by electromagnetic field, as was considered in \cite{2}. One of the aims of this article is to resolve this problem.

In a $(D+1)$-dimensional Minskowski spacetime with metric
$$ds^2=\eta_{\mu\nu}dx^{\mu}dx^{\nu}=(dx^0)^2-(dx^1)^2-\ldots (dx^D)^2,$$the field strength of  electromagnetic field is represented by the two form $F=dA=F_{\mu\nu}dx^{\mu}dx^{\nu}$.
As in \cite{2}, we introduce the potential $$A=A_{\mu}dx^{\mu}\;\left(:=\sum_{\mu=0}^DA_{\mu}dx^{\mu}\right)$$ so that $$F=dA=\left(\frac{\pa A_{\nu}}{\pa x^{\mu}}-\frac{\pa A_{\mu}}{\pa x^{\nu}}\right)dx^{\mu}dx^{\nu}.$$
The action is the standard one: \begin{equation*}S=-\frac{1}{4}\int \sqrt{|g|}F_{\mu\nu}F^{\mu\nu}d^{D+1}x,\end{equation*}where $F^{\mu\nu}=g^{\mu\eta}g^{\nu \kappa}F_{\eta\kappa}$, and the equation of motion is
\begin{equation}\label{eq01_03}
 \frac{1}{\sqrt{|g|}}\frac{\pa}{\pa x^{\nu}}\left( \sqrt{|g|}F^{\mu\nu}\right) = \frac{1}{\sqrt{|g|}}\frac{\pa}{\pa x^{\nu}} \left( \sqrt{|g|}g^{\mu\eta}g^{\nu\kappa}\left[\frac{\pa A_{\kappa}}{\pa x^{\eta}}-\frac{\pa A_{\eta}}{\pa x^{\kappa}}\right]\right) =0.
\end{equation} As is well-known, there is a gauge degree of freedom given by $$A\mapsto A+d\varphi,$$ where $\varphi$ is any function. This degree of freedom can be eliminated by imposing the radiation gauge \begin{equation}\label{eq01_04_2}A_0=0, \hspace{1cm}\frac{1}{\sqrt{|g|}}\frac{\pa}{\pa x^{\mu}}\left(\sqrt{|g|}A^{\mu}\right)=0,\hspace{1cm}A^{\mu}=g^{\mu\nu}A_{\nu}.\end{equation}
Under this gauge condition, the equation of motion \eqref{eq01_03} is equivalent to the Laplace equation
\begin{equation}\label{eq01_04_6}
\begin{split}
\Delta A_{\mu}= -\frac{g_{\mu\gamma}}{\sqrt{|g|}}\frac{\pa}{\pa x^{\nu}}\left( \sqrt{|g|}g^{\gamma\eta}g^{\nu\kappa}\left[\frac{\pa A_{\kappa}}{\pa x^{\eta}}-\frac{\pa A_{\eta}}{\pa x^{\kappa}}\right]\right)+\frac{\pa}{\pa x^{\mu}}\left(\frac{1}{\sqrt{|g|}}\frac{\pa}{\pa x^{\nu}}\left(\sqrt{|g|}g^{\nu\kappa}A_{\kappa}\right)\right)=0
\end{split}
\end{equation}on the one-form $A$. There has been a work \cite{17} which computed Casimir energy for Laplace operator acting on $p$-forms ($0\leq p \leq D$) on a $D$-dimensional ball $B^D$. The boundary conditions imposed on the boundary of the ball -- the $(D-1)$-sphere, are the so-called  absolute and  relative boundary conditions \cite{18}. In   spherical coordinates:
\begin{equation}\label{eq01_04_1}
\begin{split}\left\{\begin{aligned}
x^0=&t\\
x^1=&r\cos\theta_1\\
x^2=&r\sin\theta_1\cos\theta_2\\
&\vdots\\
x^{D-1}=&r\sin\theta_1\ldots\sin\theta_{D-2}\cos\phi\\
x^D=&r\sin\theta_1\ldots\sin\theta_{D-2}\sin\phi\end{aligned}\right.,\hspace{2cm}\left. \begin{aligned}&r\geq 0,\\
0\leq &\theta_i \leq \pi, \hspace{0.5cm} 1\leq i\leq D-2\\
0\leq & \phi=\theta_{D-1} \leq 2\pi,
\end{aligned}\right.,
\end{split}
\end{equation}the absolute boundary conditions for one-forms on a $D$-dimensional ball $B^D$ read as \cite{19, 17}:
\begin{equation}\label{eq01_03_2}
\left. \frac{\pa A_{\theta_i}}{\pa r}  \right|_{\pa B^D}=0,\hspace{1cm}\left.A_r\right|_{\pa B^D}=0;
\end{equation} and the relative boundary conditions read as \cite{19,17}:
\begin{equation}\label{eq01_03_3}
\left.  A_{\theta_i} \right|_{\pa B^D}=0,\hspace{1cm}\left.\left(\frac{\pa}{\pa r}+\frac{D-1}{r}\right)A_r\right|_{\pa B^D}=0.
\end{equation}
However, as a potential for the massless vector field, the boundary conditions imposed on $A$ as discussed in \cite{2} are completely different. In fact, the boundary conditions are imposed on the field $F=dA$ rather than the potential $A$. Generalizing the perfectly conducting  and infinitely permeable boundary conditions in $D=3$ dimension,  for any $D$, the perfectly conducting boundary condition is specified  by \cite{2}:
$$\left.n^{\mu}F^*_{\mu \nu_1\ldots\nu_{D-2}}\right|_{S}=0,$$ and the infinitely permeable condition is specified by  \cite{2}:
\begin{equation*}
\left.n^{\mu}F_{\mu\nu}\right|_{S}=0.
\end{equation*}Here $n^{\mu}$ is the spacelike vector normal to the surface $S$, and
$$F^*_{\mu_1\ldots \mu_{D-1}}=\varepsilon_{\mu_1\ldots\mu_{D-1}\nu\lambda}F^{\nu\lambda}$$ is the $(D-1)$-form dual to $F$.
These two boundary conditions are gauge invariant since $d(d\varphi)=0$.   In terms of the spherical coordinates, the perfectly conducting boundary condition on the boundary of a $D$-dimensional ball $B^D$ can be explicitly written as
\begin{equation}\label{eq01_03_4}
\begin{split}
\left.\left(\frac{\pa A_{\theta_i}}{\pa \theta_j}-\frac{\pa A_{\theta_j}}{\pa \theta_i}\right)\right|_{\pa B^D}=0,\hspace{1cm} 1\leq i<j\leq D-1,
\end{split}
\end{equation}and the infinitely permeable boundary condition can be written as
\begin{equation}\label{eq01_03_5}
\left.\left(\frac{\pa A_{\theta_i}}{\pa r}-\frac{\pa A_{r}}{\pa \theta_i}\right)\right|_{\pa B^D}=0,\hspace{1cm} 1\leq i\leq D-1.
\end{equation} At the first sight, the absolute and relative boundary conditions   \eqref{eq01_03_2} and \eqref{eq01_03_3} and the perfectly conducting and infinitely permeable boundary conditions \eqref{eq01_03_4} and \eqref{eq01_03_5} are completely different. A result we are going to show in this article is that on a $D$-dimensional ball, the absolute boundary condition is equivalent to the infinitely permeable boundary condition, and the relative boundary condition is equivalent to the perfectly conducting boundary condition. This gives a physical interpretation to the absolute and relative boundary conditions considered in \cite{17}.

After writing out explicitly the eigenmodes of the electromagnetic field in spherical coordinates, we    compute the Casimir energy using cut-off regularization and zeta function regularization. The Casimir energy for the region inside the spherical shell and the region outside the spherical shell will be computed separately. The Casimir energy for a single spherical shell configuration is obtained as a sum of the two.     In the last ten years, there has been an interest in considering double spherical shell configuration. In \cite{20}, the  stress energy tensors for a massless scalar field inside and outside a $D$-dimensional spherical shell, as well as inside the region between two spherical shells have been computed. This has been generalized to spinor fields and spacetimes with different geometries \cite{21,22,23}. In \cite{38, 41, 24}, the Casimir effect due to  electromagnetic field with perfectly conducting boundary conditions on two concentric three dimensional spherical shells   was   considered.   In this article, we are going to generalize this to $D$-dimensions.

The layout of this article is as follows. In Section \ref{s2}, we derive explicitly the eigenmodes of electromagnetic field in $D$-dimensional spherically symmetric cavity using spherical coordinates. Both perfectly conducting and infinitely permeable boundary conditions are considered. In Section \ref{s3}, we compute the Casimir energy inside a spherical cavity by cut-off and zeta regularizations. In Section \ref{s4}, we show that the Casimir energy inside an annular region bounded by two spheres can be written as a sum of three terms: the Casimir energy inside the sphere with larger radius, the term that can be interpreted as the (renormalized) Casimir energy outside the sphere with smaller radius, and the interacting term. In Section \ref{s5}, we consider the (renormalized) Casimir energy outside a spherical shell. In Section \ref{s6}, we consider the Casimir energy of a single spherical system which is the sum of the Casimir energy inside the sphere and the Casimir energy outside the sphere. In Section \ref{s7}, we consider a double spherical shell system whose Casimir energy is obtained by summing the Casimir energy inside the sphere with smaller radius, the Casimir energy in the annular region between the two spheres, and the Casimir energy outside the sphere with larger radius. Using the result of Section \ref{s4}, this can be rewritten as the sum of the Casimir energies of the two spheres, and the interacting term.  In Section \ref{s8}, we present a method to compute the asymptotic behavior of the Casimir force when the separation between the spheres is much smaller than the radii of the spheres.   A brief conclusion of the results obtained in this article is given in Section \ref{s9}.

Throughout this article, we assume that $D\geq 3$ and we use the units where $\hbar=c=1$.
\section{Eigenmodes of the Electromagnetic Field in Spherical Coordinates}\label{s2}

In this section, we will write down explicitly  the eigenmodes of the  electromagnetic fields in $(D+1)$-dimensional  spherical coordinates that satisfy the radiation gauge \eqref{eq01_04_2}. In spherical coordinates \eqref{eq01_04_1}, the metric tensor $g_{\mu\nu}$ is
\begin{equation}
ds^2=g_{\mu\nu}dx^{\mu}dx^{\nu}=dt^2-dr^2-r^2\sum_{i=1}^{D-1} \left[\prod_{j=1}^{i-1}\sin^2\theta_j\right] d\theta_i^2,
\end{equation}and $$\sqrt{|g|}=r^{D-1}\prod_{i=1}^{D-2}\sin^{D-i-1}\theta_i.$$We want to find explicitly  the one forms $$A=A_{\mu}dx^{\mu}=A_tdt+A_rdr+A_{\theta_1}d\theta_1+\ldots+A_{\theta_{D-2}}d\theta_{D-2}+A_{\phi}d\phi$$which are solutions of the equation of motion \eqref{eq01_03}
 subject to the gauge condition \eqref{eq01_04_2}. In spherical coordinates, the gauge condition  is given by
\begin{equation}\label{eq01_04_5}\begin{split}
A_t=&0, \\\mathfrak{G}=&\frac{1}{\sqrt{|g|}}\frac{\pa}{\pa x^{\mu}}\left(\sqrt{|g|}A^{\mu}\right)=-\frac{1}{r^{D-1}}\frac{\pa}{\pa r}\left( r^{D-1}A_r\right)-\frac{1}{r^2}\sum_{i=1}^{D-1}\frac{1}{\left[\prod_{j=1}^{i-1}\sin^2\theta_j \right]}\frac{1}{\sin^{D-i-1}\theta_i} \frac{\pa}{\pa\theta_i}\left(\sin^{D-i-1}\theta_iA_{\theta_i}\right)=0.\end{split}\end{equation}
In spherical coordinates, the equation  of motion \eqref{eq01_03} or the   Laplace equation \eqref{eq01_04_6}   equivalent to \eqref{eq01_03}  under the radiation gauge \eqref{eq01_04_5} are quite complicated. We will consider another set of equivalent equations. Adding $\displaystyle \frac{2}{r} \mathfrak{G}$ to the Laplace equation \eqref{eq01_04_6} when $\mu=r$, we obtain the equation
\begin{equation}\label{eq01_04_8}
\begin{split}
\frac{\pa^2}{\pa t^2}A_r-\frac{\pa^2A_r}{\pa r^2}-\frac{D+1}{r}\frac{\pa A_r}{\pa r}-\frac{D-1}{r^2}A_r-\sum_{j=1}^{D-1}\frac{1}{r^2\sin^2\theta_1\ldots\sin^2\theta_{j-1}\sin^{D-j-1}\theta_j}\frac{\pa}{\pa \theta_j}\left( \sin^{D-j-1}\theta_j\frac{\pa A_{r}}{\pa \theta_j} \right)=0.
\end{split}
\end{equation}For $1\leq i\leq D-1$, adding $\displaystyle \frac{2\cos\theta_i}{\sin\theta_i}  \mathfrak{G} $ to the Laplace equation \eqref{eq01_04_6}  when $\mu=\theta_i$, we obtain the equation
\begin{equation}\label{eq01_04_9}
\begin{split}
&\frac{\pa^2}{\pa t^2}A_{\theta_i}-\frac{1}{r^{D-3}}\frac{\pa}{\pa r}\left(r^{D-3}\frac{\pa A_{\theta_i}}{\pa r}\right) -\sum_{j=1}^{i-1}\frac{1}{r^2\sin^2\theta_1\ldots\sin^2\theta_{j-1}\sin^{D-j-3}\theta_j}\frac{\pa}{\pa\theta_j}\left(\sin^{D-j-3}\theta_j\frac{\pa A_{\theta_i}}{\pa \theta_j}\right)\\&-\frac{1}{r^2\sin^2\theta_1\ldots\sin^2\theta_{i-1}\sin^{2}\theta_i}\frac{\pa}{\pa\theta_i}\left(\frac{1}{ \sin^{D-i-3}\theta_i}
\frac{\pa}{\pa\theta_i}\left(\sin^{D-i-1}\theta_iA_{\theta_i}\right)\right)\\&-\sum_{j=i+1}^{D-1}\frac{1}{r^2\sin^2\theta_1\ldots\sin^2\theta_{j-1}\sin^{D-j-1}\theta_j}\frac{\pa}{\pa\theta_j}\left(\sin^{D-j-1}\theta_j\frac{\pa A_{\theta_i}}{\pa \theta_j}\right)\\&- \frac{2}{r}\frac{\pa A_r}{\pa\theta_i}-2\sum_{j=1}^{i-1}\frac{\cos\theta_j}{\sin\theta_j}\frac{1}{r^2\sin^2\theta_1\ldots\sin^2\theta_{j-1}}\frac{\pa A_{\theta_j} }{\pa\theta_i}\\&-
\frac{2\cos\theta_i}{\sin\theta_i} \frac{1}{r^{D-1}}\frac{\pa}{\pa r}\left(r^{D-1}A_r\right)-
\frac{2\cos\theta_i}{\sin\theta_i}\sum_{j=1}^{i-1}\frac{1}{r^2\sin^2\theta_1\ldots\sin^2\theta_{j-1}
\sin^{D-j-1}\theta_j}\frac{\pa}{\pa \theta_j}\left(\sin^{D-j-1}\theta_jA_{\theta_j}\right)=0.
\end{split}
\end{equation}Notice that \eqref{eq01_04_8} only contains $A_r$, and \eqref{eq01_04_9} only contains $A_r$ and $A_j$ with $1\leq j\leq i$. To further simplify the problem,
  we divide the energy eigenmodes of the field $F=dA$ into TE modes and TM modes as in the $D=3$ case. Since the component $E_i$ of the electric field is identified with $  F_{0i}$ and the component $B_i$ of the  magnetic field is identified with $ \{ F_{jk}\,:\, 1\leq j<k\leq D, \,j,k\neq i\}$ \cite{2},  on a $D$-dimensional ball or annular region,  TE modes and TM modes are defined as modes satisfying
\begin{equation}\label{eq01_04_4}
\text{TE modes}:\hspace{0.5cm} F_{0r}=0,\hspace{2cm}\text{TM modes:}\hspace{0.5cm}F_{\theta_j\theta_k}=0,\hspace{0.5cm}1\leq j<k\leq D-1.
\end{equation}Interestingly, we find that this division into TE and TM modes are exactly the same as the decomposition
$$A=A^T+\mathfrak{A}(\varphi)$$ for a one form $A$ satisfying the transversality condition $$\frac{1}{\sqrt{|g|}}\frac{\pa}{\pa x^{\mu}}\left(\sqrt{|g|}A^{\mu}\right)=0$$ discussed in \cite{17,18}.  To be more precise, the condition for TE modes $F_{0r}=0$ is satisfied if and only if $A_r=0$, and this is exactly the characterization of $A^T$ in \cite{17,18}. For the term $\mathfrak{A}(\varphi)$, notice that the condition on TM modes \eqref{eq01_04_4} implies that the corresponding $A$ can be written as
\begin{equation*}
A=A_rdr+e^{-i\omega t}f(r)\tilde{A},
\end{equation*}where $\tilde{A}$ is a closed one form (i.e. $d\tilde{A}=0$) on the unit $(D-1)$-dimensional sphere $S^{D-1}$. Since $D\geq 3$, the first cohomology of $S^{D-1}$ vanishes.   Therefore there exists a function $\tilde{\varphi}$ on $S^{D-1}$ such that $$d\tilde{\varphi}=\tilde{A}.$$In other words,
 for $1\leq i\leq D-1$,
$$A_{\theta_i}=e^{-i\omega t} f(r) \frac{\pa\tilde{\varphi}}{\pa\theta_i}.$$The equation of motion \eqref{eq01_03} for $\mu=\theta_i$, $1\leq i\leq D-1$ becomes
\begin{equation*}\begin{split}
0=-\frac{g_{\theta_i\theta_i}}{\sqrt{|g|}}\frac{\pa}{\pa t}\left(\sqrt{|g|}g^{\theta_i\theta_i}g^{tt}F_{\theta_i t}\right)-\frac{g_{\theta_i\theta_i}}{\sqrt{|g|}}\frac{\pa}{\pa r}\left(\sqrt{|g|}g^{\theta_i\theta_i}g^{rr}F_{\theta_i r}\right)=
\frac{\pa^2  A_{\theta_i}}{\pa t^2} +\frac{1}{r^{D-3}}
\frac{\pa}{\pa r}\left(r^{D-3}\left[\frac{\pa A_r}{\pa\theta_i}-\frac{\pa A_{\theta_i}}{\pa r}\right]\right).
\end{split}\end{equation*} This implies that $$A_r=e^{-i\omega t}g(r)\tilde{\varphi},$$ where \begin{equation}\label{eq01_04_11}\frac{1}{r^{D-3}}
\frac{d}{d r}\left(r^{D-3}g(r)\right)=\frac{1}{r^{D-3}}
\frac{d}{d r}\left(r^{D-3}\frac{df}{dr}\right)+\omega^2f(r).\end{equation} Since the Laplace operator on scalar functions on the $(D-1)$-dimensional unit sphere $S^{D-1}$ is
\begin{equation*}
\Delta^{(0)}_{S^{D-1}}=-\sum_{i=1}^{D-1}\frac{1}{\left[\prod_{j=1}^{i-1}\sin^2\theta_j \right]}\frac{1}{\sin^{D-i-1}\theta_i} \frac{\pa}{\pa\theta_i}\left(\sin^{D-i-1}\theta_i\frac{\pa}{\pa \theta_i}\right),
\end{equation*} the equation for $A_r$ \eqref{eq01_04_8} and the gauge condition \eqref{eq01_04_5} imply that
\begin{equation*}
\begin{split}
-\omega^2 g(r)\tilde{\varphi}-\frac{d^2 g(r)}{d r^2}\tilde{\varphi}-\frac{D+1}{r}\frac{d g(r)}{d r}\tilde{\varphi}-\frac{D-1}{r^2 }g(r)\tilde{\varphi}+\frac{g(r)}{r^2}\Delta^{(0)}_{S^{D-1}}\tilde{\varphi}=0,\\
\frac{1}{r^{D-3}}\frac{d}{dr}\left(r^{D-1} g(r)\right) \tilde{\varphi} -f(r)\Delta^{(0)}_{S^{D-1}}\tilde{\varphi}=0.\end{split}
\end{equation*}These are satisfied if and only if
\begin{equation}\label{eq01_04_10}
\begin{split}
\Delta^{(0)}_{S^{D-1}}\tilde{\varphi}=\lambda^2\tilde{\varphi},\\
\frac{d^2 g(r)}{d r^2}+\frac{D+1}{r}\frac{d g(r)}{d r}-\frac{\lambda^2-(D-1)}{r^2 }g(r)+\omega^2g(r)=0,\\
\frac{1}{r^{D-3}}\frac{d}{dr}\left(r^{D-1} g(r)\right)=\lambda^2 f(r).
\end{split}
\end{equation}The first equation implies that $\tilde{\varphi}$ is an eigenfunction of the Laplace operator on the unit sphere $S^{D-1}$, with eigenvalue $\lambda^2$. It is well known that the complete set of eigenfunctions on $S^{D-1}$ is
\begin{equation*}\begin{split}
&Y_{\boldsymbol{m}}(\boldsymbol{\theta})=\left(\prod_{i=1}^{D-2} P_{m_i;D-i-1}^{(m_{i+1})} (\cos\theta_i)\right)e^{im_{D-1}\phi},\\
-m_{D-2}\leq &m_{D-1}\leq m_{D-2},\;\; 0\leq m_{D-2}\leq m_{D-3}\leq \ldots\leq m_1.
\end{split}\end{equation*}Here
$$P_{m_i;D-i-1}^{(0)}(x)=C_{m_i}^{\frac{D-i-1}{2}}(x)$$ is the Gegenbauer polynomial of degree $m_i$ and order $(D-i-1)/2$ defined by the generating function \cite{26}:
\begin{equation*}
\frac{1}{(1-2xw+w^2)^{\frac{D-i-1}{2}}}=\sum_{j=0}^{\infty} C_{j}^{\frac{D-i-1}{2}}(x) w^j,
\end{equation*} and
\begin{equation*}
P_{m_i;D-i-1}^{(m_{i+1})} (x)=(1-x^2)^{\frac{|m_{i+1}|}{2}}C_{m_i-|m_{i+1}|}^{\frac{D-i-1}{2}+|m_{i+1}|}(x).
\end{equation*}The   eigenvalue of $Y_{\boldsymbol{m}}(\boldsymbol{\theta})$ is $$\lambda^2=l(l+D-2),\hspace{1cm} l:=m_1.$$For a given $l\geq 0$, the degeneracy of the  eigenvalue $\lambda^2=l(l+D-2)$ is equal to
\begin{equation}\label{eq01_04_14}
b_D(l) =  \sum_{m_2=0}^{l}\sum_{m_3=0}^{m_2}\ldots\sum_{m_{D-2}=0}^{m_{D-3}}\sum_{m_{D-1}=-m_{D-2}}^{m_{D-2}}1=\frac{(2l+D-2)(l+D-3)!}{(D-2)! l!}.
\end{equation}With $\lambda^2=l(l+D-2)$, the equation for $g(r)$ \eqref{eq01_04_10} becomes
\begin{equation*}
\begin{split}
\frac{d^2 g(r)}{d r^2}+\frac{D+1}{r}\frac{d g(r)}{d r}-\frac{l(l+D-2)-(D-1)}{r^2 }g(r)+\omega^2g(r)=0,
\end{split}
\end{equation*}whose general solution is given by
\begin{equation}\label{eq01_04_12}
g(r)= r^{-\frac{D}{2}}\left(C_1J_{l+\frac{D-2}{2}}(\omega r)+C_2N_{l+\frac{D-2}{2}}(\omega r)\right),
\end{equation}where $J_{\nu}(z)$ and $N_{\nu}(z)$ are Bessel functions of first and second kind. The third equation in \eqref{eq01_04_10} implies that $l\neq 0$ and
\begin{equation}\label{eq01_04_13}
f(r)=\frac{1}{l(l+D-2)}\frac{1}{r^{D-3}}\frac{d}{dr}\left(r^{D-1} g(r)\right)=\frac{1}{l(l+D-2)}\frac{1}{r^{D-3}}\frac{d}{dr}\left(r^{\frac{D-2}{2}}
\left[C_1J_{l+\frac{D-2}{2}}(\omega r)+C_2N_{l+\frac{D-2}{2}}(\omega r)\right]\right).
\end{equation}One can check that $g(r)$ and $f(r)$ given by \eqref{eq01_04_12} and \eqref{eq01_04_13} satisfy the equation \eqref{eq01_04_11}. As a summary, the set of TM modes are given by $A_t=0$,
\begin{equation}\label{eq3_3_1}
\begin{split}
\left\{\begin{aligned} A_r=&l(l+D-2)e^{-i\omega t} r^{-\frac{D}{2}}\left(C_1J_{l+\frac{D-2}{2}}(\omega r)+C_2N_{l+\frac{D-2}{2}}(\omega r)\right)\left(\prod_{j=1}^{D-2} P_{m_j;D-j-1}^{(m_{j+1})} (\cos\theta_j)\right)e^{im_{D-1}\phi}\\
A_{\theta_i}=& \frac{e^{-i\omega t}}{r^{D-3}}\frac{d}{dr}\left(r^{\frac{D-2}{2}}
\left[C_1J_{l+\frac{D-2}{2}}(\omega r)+C_2N_{l+\frac{D-2}{2}}(\omega r)\right]\right)\frac{d}{d\theta_i}\left(\left[\prod_{j=1}^{D-2}  P_{m_j;D-j-1}^{(m_{j+1})} (\cos\theta_j) \right]e^{im_{D-1}\phi}\right), \hspace{0.5cm}1\leq i\leq D-1\end{aligned}\right.\\
-m_{D-2}\leq m_{D-1}\leq m_{D-2},\;\; 0\leq m_{D-2}\leq m_{D-3}\leq \ldots\leq m_1, \;\; m_1\geq 1.\hspace{5cm}
\end{split}
\end{equation}It can be written as $\mathfrak{A}(\varphi)$, where $$\varphi=e^{-i\omega t} r^{-\frac{D-2}{2}}\left(C_1J_{l+\frac{D-2}{2}}(\omega r)+C_2N_{l+\frac{D-2}{2}}(\omega r)\right)\left(\prod_{i=1}^{D-2} P_{m_i;D-i-1}^{(m_{i+1})} (\cos\theta_i)\right)e^{im_{D-1}\phi}$$is an eigenfunction of the Laplace operator, and
\begin{equation*}
\begin{split}
\mathfrak{A}(\varphi)_r=&\Delta^{(0)}_{S^{D-1}}\left(r^{-1}\varphi\right)\\
\mathfrak{A}(\varphi)_{\theta_i}=&r\left(\frac{\pa}{\pa r}+\frac{D-2}{r}\right)\frac{\pa \varphi}{\pa\theta_i},
\end{split}
\end{equation*}as defined in \cite{17,18}. For fixed $l\geq 1$, the number of TM modes with $m_1=l$ is equal to $b_D(l)$ \eqref{eq01_04_14}.

Next, we turn to explicit solutions of the TE modes. In this case, $A_r=0$ and $A=e^{-i\omega t}f(r) \tilde{A}$ for some transversal one form $\tilde{A}$ on the unit sphere. The set of transversal one forms on a unit sphere $S^{D-1}$ has been constructed in \cite{27,28,29} using more sophisticated language and the results have been employed in \cite{17,18}.  In fact, we can directly  solve  the system of equations \eqref{eq01_04_8} and \eqref{eq01_04_9} for $\{A_r, A_{\theta_i}, 1\leq i \leq D-1\}$. The form of the equations \eqref{eq01_04_8} and \eqref{eq01_04_9} allows us to solve for $A_r, A_{\theta_1}, \ldots, A_{\theta_{D-2}}, A_{\phi}$ one by one using separation of variables. The technique is similar to solving the eigenvalue equation for Laplace equations on functions in spherical coordinates. Using the fact that
\begin{equation*}
\frac{d^2P_{m_i;D-i-1}^{(m_{i+1})}(\cos\theta_i)}{d\theta_i^2}+(D-i-1)\frac{\cos\theta_i}{\sin\theta_i}\frac{dP_{m_i;D-i-1}^{(m_{i+1})}(\cos\theta_i)}{d\theta_i}+
\left(m_i(m_i+D-i-1)-\frac{m_{i+1}(m_{i+1}+D-i-2)}{\sin^2\theta_i}\right)=0,
\end{equation*}one can show that the   TE modes can be divided into $(D-2)$ sets corresponding to   $(D-2)$ polarizations, where for $1\leq i\leq D-3$, the $i^{\text{th}}$ set is given by
\begin{equation}\label{eq3_3_2}
\begin{split}
\left\{\begin{aligned}A_t=&A_r=A_{\theta_1}=\ldots=A_{\theta_{i-1}}=0\\
A_{\theta_i}=&m_{i+1}(m_{i+1}+D-i-2)e^{-i\omega t}f(r)\prod_{j=1}^{i-1}\left(\sin\theta_j P_{m_j;D-j-1}^{(m_{j+1})}(\cos\theta_j)\right)\frac{P_{m_i;D-i-1}^{(m_{i+1})}(\cos\theta_i)}{\sin\theta_i}\prod_{j=i+1}^{D-2}
P_{m_j;D-j-1}^{(m_{j+1})}(\cos\theta_j)e^{im_{D-1}\phi}\\
A_{\theta_l}=& e^{-i\omega t}f(r)\prod_{j=1}^{i-1}\left(\sin\theta_jP_{m_j;D-j-1}^{(m_{j+1})}(\cos\theta_j)\right)\frac{1}{\sin^{D-i-3}\theta_i}
\frac{d}{d\theta_i}\left(\sin^{D-i-2}\theta_i P_{m_i;D-i-1}^{(m_{i+1})}(\cos\theta_i)\right)\\&\times\frac{d}{d\theta_l}\left(\prod_{j=i+1}^{D-2}P_{m_j;D-j-1}^{(m_{j+1})}
(\cos\theta_j)e^{im_{D-1}\phi}\right),\hspace{1cm}i+1\leq l\leq D-1,\end{aligned}\right.\\
f(r)=r^{\frac{4-D}{2}}\left(C_1J_{l+\frac{D-2}{2}}(\omega r)+C_2N_{l+\frac{D-2}{2}}(\omega r)\right),\hspace{6cm}\\
-m_{D-1}\leq m_{D-2}\leq m_{D-1},\hspace{0.5cm} 0\leq m_{D-2}\leq \ldots \leq m_1,\hspace{0.5cm}m_{i+1}\geq 1,\hspace{5cm}
\end{split}
\end{equation}and the $(D-2)^{\text{th}}$ set is
\begin{equation}\label{eq3_3_3}
\begin{split}
\left\{\begin{aligned}A_t=&=A_r=A_{\theta_1}=\ldots=A_{\theta_{D-3}}=0\\
A_{\theta_{D-2}}=&m_{D-1} e^{-i\omega t}f(r)\prod_{j=1}^{D-3}\left(\sin\theta_j P_{m_j;D-j}^{(m_{j+1})}(\cos\theta_j)\right)\frac{P_{m_{D-2};1}^{(m_{D-1})}(\cos\theta_{D-2})}{\sin\theta_{D-2}} e^{im_{D-1}\phi}\\
A_{\phi}=&i e^{-i\omega t}f(r)\prod_{j=1}^{D-3}\left(\sin\theta_jP_{m_j;D-j-1}^{(m_{j+1})}(\cos\theta_j)\right) \sin \theta_{D-2}
\frac{d}{d\theta_{D-2}}\left(   P_{m_{D-2};1}^{m_{D-1}}(\cos\theta_{D-2})\right) e^{im_{D-1}\phi}\end{aligned}\right.\\
f(r)=r^{\frac{4-D}{2}}\left(C_1J_{l+\frac{D-2}{2}}(\omega r)+C_2N_{l+\frac{D-2}{2}}(\omega r)\right),\hspace{6cm}\\
-m_{D-1}\leq m_{D-2}\leq m_{D-1},\hspace{0.5cm} 1\leq m_{D-2}\leq \ldots \leq m_1.\hspace{5cm}
\end{split}
\end{equation}  Observe that for $1\leq i\leq D-3$, the $i^{\text{th}}$ set can be considered as constructed from the transversal one forms on $S^{D-i}$ which satisfies $A_{\theta_i}\neq 0$. For fixed $l\geq 1$, the number of TE modes with $m_1=l$ is given by
\begin{equation*}
h_D(l)=\sum_{i=1}^{D-2}h_{D;i}(l),
\end{equation*}where $h_{D;i}(l)$ is the number of modes in the $i^{\text{th}}$-set with $m_1=l$, given by
\begin{equation*}
\begin{split}
h_{D;i}(l)=&\sum_{m_2=1}^{l}\sum_{m_3=1}^{m_2}\ldots\sum_{m_{i+1}=1}^{m_i}\sum_{m_{i+2}=0}^{m_{i+1}}\ldots\sum_{m_{D-2}=0}^{m_{D-3}}\sum_{m_{D-1}=-m_{D-2}}^{m_{D-2}}1,\hspace{1cm}1\leq i\leq D-3,\\
h_{D;D-2}(l)=&\sum_{m_2=1}^{l}\sum_{m_3=1}^{m_2}\ldots \sum_{m_{D-3}=1}^{m_{D-4}} \sum_{m_{D-2}=1}^{m_{D-3}}\sum_{m_{D-1}=-m_{D-2}}^{m_{D-2}}1.
\end{split}
\end{equation*}It is easy to see that
\begin{equation*}
h_D(l)=h_{D,1}(l)+\sum_{m_2=1}^l h_{D-1}(m_2)=\sum_{m_2=1}^l \left(b_{D-1}(m_2)+h_{D-1}(m_2)\right)=\sum_{m_2=1}^l\left( \frac{(2m_2+D-3)(m_2+D-4)!}{(D-3)!m_2!}+h_{D-1}(m_2)\right).
\end{equation*}Using this, one can show by induction on $D$ and $l$ that
$$h_D(l)= \frac{l(l+D-2)(2l+D-2)(l+D-4)!}{(D-3)!(l+1)!},$$ in agreement with the result in \cite{27}.
Since we find $(D-2)$ polarizations for the TE modes, and one polarization for TM modes, we have altogether $(D-1)$ polarizations as expected for electromagnetic field in $D$-dimensional space.

\bigskip
\begin{table} \caption{\label{t1}Eigenmodes of electromagnetic field I: Perfectly conducting boundary condition}

\vspace{0.3cm}
\begin{tabular}{ c c c}
\hline
\hline

\vspace{0.3cm}
 & TE modes & TM modes\\

\vspace{0.3cm}
&Inside a sphere with radius $a$ & \\

 \vspace{0.3cm}
Eigenmodes & $\omega=\omega_{lj}^{\text{TE}}, l,j=1,2,\ldots$ solution  of& $\omega=\omega^{\text{TM}}_{lj}, l,j=1, 2, \ldots$ solution of \\

\vspace{0.3cm}
 & $\displaystyle  J_{l+\frac{D-2}{2}}\left(\omega  a\right)=0$ & $\displaystyle \left.\frac{d}{dr}\left(r^{\frac{D-2}{2}}
 J_{l+\frac{D-2}{2}}\left(\omega  r\right) \right)\right|_{r=a}=0$\\

\vspace{0.3cm}
&Between two spheres with radius $a<b$ \\

 \vspace{0.3cm}
Eigenmodes & $\omega=\omega_{lj}^{\text{TE}}, l,j=1,2,\ldots$ solution  of& $\omega=\omega^{\text{TM}}_{lj}, l,j=1, 2, \ldots$ solution of \\

\vspace{0.3cm}
 & $\displaystyle \begin{aligned}&J_{l+\frac{D-2}{2}}(\omega a)N_{l+\frac{D-2}{2}}(\omega b)\\-&J_{l+\frac{D-2}{2}}(\omega b)N_{l+\frac{D-2}{2}}(\omega a)=0\end{aligned}$ & $\displaystyle \begin{aligned}&\left.\frac{d}{dr}\left(r^{\frac{D-2}{2}}
 J_{l+\frac{D-2}{2}}(\omega r) \right)\right|_{r=a}\left.\frac{d}{dr}\left(r^{\frac{D-2}{2}}
 N_{l+\frac{D-2}{2}}(\omega r) \right)\right|_{r=b}\\& -\left.\frac{d}{dr}\left(r^{\frac{D-2}{2}}
 J_{l+\frac{D-2}{2}}(\omega r) \right)\right|_{r=b}\left.\frac{d}{dr}\left(r^{\frac{D-2}{2}}
 N_{l+\frac{D-2}{2}}(\omega r) \right)\right|_{r=a}=0\end{aligned}$\\

\vspace{0.3cm}
Degeneracy & $\displaystyle h_D(l)= \frac{l(l+D-2)(2l+D-2)(l+D-4)!}{(D-3)!(l+1)!}$  & $\displaystyle b_D(l) =  \frac{(2l+D-2)(l+D-3)!}{(D-2)! l!}$\\

\hline
\hline
\end{tabular}
\end{table}

\vspace{0.3cm}
\begin{table} \caption{\label{t1_2}Eigenmodes of electromagnetic field II: Infinitely permeable boundary condition}

\vspace{0.3cm}
\begin{tabular}{ c c c}
\hline
\hline

\vspace{0.3cm}
 & TE modes & TM modes\\

\vspace{0.3cm}
&Inside a sphere with radius $a$ & \\

 \vspace{0.3cm}
Eigenmodes & $\omega=\omega_{lj}^{\text{TE}}, l,j=1,2,\ldots$ solution  of& $\omega=\omega^{\text{TM}}_{lj}, l,j=1, 2, \ldots$ solution of \\

\vspace{0.3cm}
 &$\displaystyle \left.\frac{d}{dr}\left(r^{\frac{4-D}{2}}
 J_{l+\frac{D-2}{2}}\left(\omega  r\right) \right)\right|_{r=a}=0$ & $\displaystyle  J_{l+\frac{D-2}{2}}\left(\omega  a\right)=0$   \\

\vspace{0.3cm}
&Between two spheres with radius $a<b$ \\

 \vspace{0.3cm}
Eigenmodes & $\omega=\omega_{lj}^{\text{TE}}, l,j=1,2,\ldots$ solution  of& $\omega=\omega^{\text{TM}}_{lj}, l,j=1, 2, \ldots$ solution of \\

\vspace{0.3cm}
 &  $\displaystyle \begin{aligned}&\left.\frac{d}{dr}\left(r^{\frac{4-D}{2}}
 J_{l+\frac{D-2}{2}}(\omega r) \right)\right|_{r=a}\left.\frac{d}{dr}\left(r^{\frac{4-D}{2}}
 N_{l+\frac{D-2}{2}}(\omega r) \right)\right|_{r=b}\\& -\left.\frac{d}{dr}\left(r^{\frac{4-D}{2}}
 J_{l+\frac{D-2}{2}}(\omega r) \right)\right|_{r=b}\left.\frac{d}{dr}\left(r^{\frac{4-D}{2}}
 N_{l+\frac{D-2}{2}}(\omega r) \right)\right|_{r=a}=0\end{aligned}$ &$\displaystyle \begin{aligned}&J_{l+\frac{D-2}{2}}(\omega a)N_{l+\frac{D-2}{2}}(\omega b)\\-&J_{l+\frac{D-2}{2}}(\omega b)N_{l+\frac{D-2}{2}}(\omega a)=0\end{aligned}$  \\

\vspace{0.3cm}
Degeneracy & $\displaystyle h_D(l)= \frac{l(l+D-2)(2l+D-2)(l+D-4)!}{(D-3)!(l+1)!}$  & $\displaystyle b_D(l) =  \frac{(2l+D-2)(l+D-3)!}{(D-2)! l!}$\\

\hline
\hline
\end{tabular}
\end{table}

 Next we turn to the questions of imposing boundary conditions. In the case the electromagnetic field is confined in a $D$-dimensional spherical cavity of radius $a$, the regularity of the field at the origin $r=0$ requires that the coefficient $C_2$ in the field modes \eqref{eq3_3_1}, \eqref{eq3_3_2} and \eqref{eq3_3_3} vanishes.  The perfectly conducting boundary  condition \eqref{eq01_03_4} is then satisfied if and only if   for the TE modes,
\begin{equation*}
 J_{l+\frac{D-2}{2}}(\omega a)=0,
\end{equation*}and for the TM modes,
\begin{equation*}
\left.\frac{d}{dr}\left(r^{\frac{D-2}{2}}
 J_{l+\frac{D-2}{2}}(\omega r) \right)\right|_{r=a}=0.
\end{equation*}One can check directly that these coincide  with the relative boundary conditions \eqref{eq01_03_3}.
On the other hand,
 the infinitely permeable conditions \eqref{eq01_03_5} are satisfied if and only if   for the TE modes,
\begin{equation*}
\left.\frac{d}{dr}\left(r^{\frac{4-D}{2}} J_{l+\frac{D-2}{2}}(\omega r)\right)\right|_{r=a}=0,
\end{equation*}and for the TM modes,
\begin{equation*}
 J_{l+\frac{D-2}{2}}(\omega a)  =0.
\end{equation*}One can check directly that these coincide  with the absolute boundary conditions \eqref{eq01_03_2}. Thus the Casimir energies computed in \cite{17} can be interpreted as the Casimir energies in the interior of a $D$-dimensional spherical cavity with perfectly conducting boundary conditions on the spherical boundary (corresponding to relative boundary conditions) or with infinitely permeable boundary conditions on the spherical boundary (corresponding to the absolute boundary conditions).

For electromagnetic fields confined between two concentric spheres with radii $a$ and $b$, where $0<a<b$,   the boundary conditions on the surface $r=a$ and on the surface $r=b$ give rise to a system of equations satisfied by $C_1, C_2$. For perfectly conducting conditions on both shells, we have
\begin{equation*}
\begin{split}
&\text{TE modes:}\hspace{1cm}\left\{ \begin{aligned} C_1J_{l+\frac{D-2}{2}}(\omega a)+C_2N_{l+\frac{D-2}{2}}(\omega a)=0 \\
C_1J_{l+\frac{D-2}{2}}(\omega b)+C_2N_{l+\frac{D-2}{2}}(\omega b)=0\end{aligned}\right.\\
&\text{TM modes:}\hspace{1cm}\left\{ \begin{aligned}
\left.\frac{d}{dr}\left(r^{\frac{D-2}{2}}
 \left[C_1J_{l+\frac{D-2}{2}}(\omega r)+C_2N_{l+\frac{D-2}{2}}(\omega r) \right] \right)\right|_{r=a}=0,\\
 \left.\frac{d}{dr}\left(r^{\frac{D-2}{2}}
 \left[C_1J_{l+\frac{D-2}{2}}(\omega r)+C_2N_{l+\frac{D-2}{2}}(\omega r) \right] \right)\right|_{r=b}=0
\end{aligned}\right..
\end{split}\end{equation*}
 The TE modes and TM modes are those $\omega$ which give rise to nontrivial solutions of $C_1, C_2$, i.e., they satisfy
\begin{align*}
&\text{TE modes:}\hspace{1cm}J_{l+\frac{D-2}{2}}(\omega a)N_{l+\frac{D-2}{2}}(\omega b)-J_{l+\frac{D-2}{2}}(\omega b)N_{l+\frac{D-2}{2}}(\omega a)=0,\\
&\text{TM modes:}\hspace{1cm}\left.\frac{d}{dr}\left(r^{\frac{D-2}{2}}
 J_{l+\frac{D-2}{2}}(\omega r) \right)\right|_{r=a}\left.\frac{d}{dr}\left(r^{\frac{D-2}{2}}
 N_{l+\frac{D-2}{2}}(\omega r) \right)\right|_{r=b}\\&\hspace{2.5cm}-\left.\frac{d}{dr}\left(r^{\frac{D-2}{2}}
 J_{l+\frac{D-2}{2}}(\omega r) \right)\right|_{r=b}\left.\frac{d}{dr}\left(r^{\frac{D-2}{2}}
 N_{l+\frac{D-2}{2}}(\omega r) \right)\right|_{r=a}=0.
\end{align*}Similarly, for infinitely permeable boundary conditions on both shells, the TE and TM modes are solutions of the following equations:
\begin{align*}
&\text{TE modes:}\hspace{1cm}\left.\frac{d}{dr}\left(r^{\frac{4-D}{2}}
 J_{l+\frac{D-2}{2}}(\omega r) \right)\right|_{r=a}\left.\frac{d}{dr}\left(r^{\frac{4-D}{2}}
 N_{l+\frac{D-2}{2}}(\omega r) \right)\right|_{r=b}\\&\hspace{2.5cm}-\left.\frac{d}{dr}\left(r^{\frac{4-D}{2}}
 J_{l+\frac{D-2}{2}}(\omega r) \right)\right|_{r=b}\left.\frac{d}{dr}\left(r^{\frac{4-D}{2}}
 N_{l+\frac{D-2}{2}}(\omega r) \right)\right|_{r=a}=0,\\
&\text{TM modes:}\hspace{1cm}J_{l+\frac{D-2}{2}}(\omega a)N_{l+\frac{D-2}{2}}(\omega b)-J_{l+\frac{D-2}{2}}(\omega b)N_{l+\frac{D-2}{2}}(\omega a)=0.
\end{align*}


The results of this section are summarized in Table \ref{t1} and Table \ref{t1_2}. Notice that when $D=3$, $h_3(l)=b_3(l)=2l+1$, the TE modes for perfectly conducting boundary conditions is the same as the TM modes for infinitely permeable boundary conditions, and the TM modes for perfectly conducting boundary conditions is the same as the TE modes for infinitely permeable boundary conditions. Therefore   when $D=3$, the Casimir energy for perfectly conducting boundary condition is the same as the Casimir energy for infinitely permeable boundary condition. This duality does not hold for $D>3$.

\section{Casimir Energy Inside a Spherical Shell}\label{s3}
\begin{table} \caption{\label{t2}Heat kernel coefficients for the interior of a spherical shell of radius $a$ with perfectly conducting boundary condition}

\begin{tabular}{c|ccccccc }
\hline
\hline
& $D=3$ & $D=4$ & $D=5$ & $D=6$ &  $D=7$ & $D=8$ & $D=9$  \\
\hline
&&&&&&\\
$c_0^{\text{int}}$ & $\displaystyle \frac{a^3}{3\sqrt{\pi}}$ &$\displaystyle \frac{3a^4}{32}$ &$\displaystyle \frac{a^5}{15\sqrt{\pi}}$&$\displaystyle \frac{5a^6}{384}$&$\displaystyle \frac{a^7}{ 140\sqrt{\pi}}$&$\displaystyle \frac{7a^8}{6144}$ & $\displaystyle \frac{a^9}{1890\sqrt{\pi}}$   \\
&&&&&&\\
$c_1^{\text{int}}$ & $0$ & $\displaystyle -\frac{\sqrt{\pi}a^3}{16}$ &$\displaystyle -\frac{a^4}{12}$ &$\displaystyle -\frac{3\sqrt{\pi}a^5}{128}$ &$\displaystyle -\frac{a^6}{60} $ &$\displaystyle -\frac{5\sqrt{\pi}a^7}{1536}$ & $\displaystyle -\frac{a^8}{560}$ \\
&&&&&&\\
$c_2^{\text{int}}$ & $\displaystyle -\frac{4a}{3\sqrt{\pi}}$&$\displaystyle -\frac{3a^2}{8}$&$\displaystyle -\frac{2a^3}{9\sqrt{\pi}}$&$\displaystyle -\frac{5a^4}{192}$&$\displaystyle 0 $&$\displaystyle  \frac{7a^6}{2304}$ & $\displaystyle \frac{a^7}{315\sqrt{\pi}}$ \\
&&&&&&\\
$c_3^{\text{int}}$ &$\displaystyle \frac{5}{8}$&$\displaystyle \frac{211\sqrt{\pi} a}{512}  $&$\displaystyle \frac{a^2}{2}$&$\displaystyle \frac{585\sqrt{\pi}a^3}{4096}$&$\displaystyle \frac{5a^4}{48}$&$\displaystyle \frac{1015\sqrt{\pi}a^5}{49152} $ & $\displaystyle  \frac{19a^6}{1680}$     \\
&&&&&& \\
$c_4^{\text{int}}$ &$\displaystyle-\frac{16}{315\sqrt{\pi}a} $&$\displaystyle -\frac{7}{20}$&$\displaystyle -\frac{796a}{945\sqrt{\pi}}$&$\displaystyle -\frac{3 a^2}{8}$&$\displaystyle -\frac{608 a^3}{1575\sqrt{\pi}} $  &$\displaystyle-\frac{197a^4}{1920}$ & $\displaystyle -\frac{346a^5}{4725\sqrt{\pi}}$\\
&&&&&& \\
$c_5^{\text{int}}$ & &$\displaystyle  \frac{1631\sqrt{\pi}}{65536a} $&$\displaystyle \frac{541}{2880} $&$\displaystyle \frac{75361\sqrt{\pi}a}{524288} $&$\displaystyle \frac{163a^2}{768} $&$\displaystyle \frac{6994813\sqrt{\pi}a^3}{94371840} $ & $ \displaystyle \frac{887a^4}{13440}$\\
&&&&&&\\
$c_6^{\text{int}}$ & &&$\displaystyle -\frac{6632}{45045\sqrt{\pi}a} $&$\displaystyle -\frac{1109}{7560 }$&$\displaystyle  -\frac{17468 a}{61425\sqrt{\pi}}$&$\displaystyle-\frac{a^2}{8} $  & $ \displaystyle -\frac{1915624 a^3}{14189175\sqrt{\pi}}$ \\
&&&&\\
$c_7^{\text{int}}$ &&&& $\displaystyle \frac{1052991 \sqrt{\pi}}{16777216 a} $&$\displaystyle \frac{143263}{967680}$&$\displaystyle \frac{231850177\sqrt{\pi}a}{3019898880} $ & $\displaystyle \frac{491a^2}{5120}$ \\
&&&&\\
$c_8^{\text{int}}$ &&&&&$\displaystyle-\frac{16438144}{72747675\sqrt{\pi}a}$&$\displaystyle -\frac{33521}{226800}$  & $\displaystyle   -0.12483a$\\
&&&&\\
$c_9^{\text{int}}$ &&&&&&$\displaystyle \frac{2580976217486942940567943799\sqrt{\pi}}{32674585544991625157478973440 a}$ & $\displaystyle 0.14319$ \\
&&&&\\
$c_{10}^{\text{int}}$ &&&&&&& $\displaystyle -\frac{0.15150}{a}$\\
&&&&\\
\hline
\hline
\end{tabular}

\end{table}

\begin{table}[h]\caption{\label{t4}Heat kernel coefficients for the interior of a spherical shell of radius $a$ with infinitely permeable boundary condition}

\begin{tabular}{c|ccccccc }
\hline
\hline
& $D=3$ & $D=4$ & $D=5$ & $D=6$ &  $D=7$ & $D=8$   & $D=9$\\
\hline
&&&&&&\\
$c_0^{\text{int}}$ & $\displaystyle \frac{a^3}{3\sqrt{\pi}}$ &$\displaystyle \frac{3a^4}{32} $ &$\displaystyle \frac{a^5}{15\sqrt{\pi}}$&$\displaystyle\frac{5a^6}{384}  $&$\displaystyle  \frac{a^7}{140\sqrt{\pi}}$&$\displaystyle  \frac{7a^8}{6144}$  &$\displaystyle   \frac{a^9}{1890\sqrt{\pi}} $   \\
&&&&&&\\
$c_1^{\text{int}}$ & $0$ & $\displaystyle \frac{\sqrt{\pi}a^3}{16} $ &$\displaystyle \frac{a^4}{12} $ &$\displaystyle \frac{3\sqrt{\pi}a^5}{128} $ &$\displaystyle \frac{a^6}{60} $ &$\displaystyle  \frac{5\sqrt{\pi}a^7}{1536}$ &$\displaystyle \frac{a^8}{560} $ \\
&&&&&&\\
$c_2^{\text{int}}$ & $\displaystyle -\frac{4a}{3\sqrt{\pi}}$&$\displaystyle -\frac{3a^2}{8} $&$\displaystyle -\frac{2a^3}{9\sqrt{\pi}} $&$\displaystyle -\frac{5a^4}{192} $&$\displaystyle  0 $&$\displaystyle \frac{7a^6}{2304} $ &$\displaystyle \frac{a^7}{315\sqrt{\pi}} $  \\
&&&&&&\\
$c_3^{\text{int}}$ &$\displaystyle\frac{5}{8}  $&$\displaystyle    -\frac{121\sqrt{\pi}}{512} a $&$ \displaystyle -\frac{a^2}{4}$&$\displaystyle-\frac{235\sqrt{\pi}a^3}{4096} $&$ \displaystyle-\frac{7a^4}{240}$&$\displaystyle -\frac{133\sqrt{\pi}a^5}{49152}  $    &$\displaystyle \frac{a^6}{1680} $   \\
&&&&&& \\
$c_4^{\text{int}}$ &$\displaystyle  -\frac{16}{315\sqrt{\pi}a}$&$\displaystyle  \frac{49}{60}$&$\displaystyle-\frac{68a}{189\sqrt{\pi}} $&$\displaystyle -\frac{a^2}{8}$&$\displaystyle -\frac{16a^3}{175\sqrt{\pi}}$  &$\displaystyle -\frac{9a^4}{640}$  &$\displaystyle -\frac{82 a^5}{33075\sqrt{\pi}} $ \\
&&&&&& \\
$c_5^{\text{int}}$ & &$\displaystyle   - \frac{2713\sqrt{\pi}}{65536a}  $&$\displaystyle \frac{2429}{2880}   $&$\displaystyle  - \frac{44071\sqrt{\pi}a}{524288}$&$\displaystyle  -\frac{23a^2}{256} $&$\displaystyle  -\frac{2036587\sqrt{\pi}a^3}{94371840} $  &$\displaystyle  -\frac{51a^4}{4480}$ \\
&&&&&&\\
$c_6^{\text{int}}$ & &&$\displaystyle -\frac{11048}{45045\sqrt{\pi}a} $&$\displaystyle \frac{6199}{7560} $&$\displaystyle -\frac{58012a}{225225\sqrt{\pi}} $&$\displaystyle -\frac{a^2}{12} $   &$\displaystyle -\frac{905896 a^3}{14189175\sqrt{\pi}} $  \\
&&&&\\
$c_7^{\text{int}}$ &&&& $\displaystyle -\frac{871339\sqrt{\pi}}{8388608a} $&$\displaystyle \frac{785567}{967680} $&$\displaystyle - \frac{28775291\sqrt{\pi}a}{377487360}
 $ &$\displaystyle  -\frac{75a^2}{1024}$  \\
&&&&\\
$c_8^{\text{int}}$ &&&&&$\displaystyle -\frac{2955168}{8083075\sqrt{\pi}a}$&$\displaystyle \frac{185449}{226800}$  &$\displaystyle -0.12063a $ \\
&&&&\\
$c_9^{\text{int}}$ &&&&&&$\displaystyle  -\frac{4030878578159017023021286121\sqrt{\pi}}{32674585544991625157478973440a}$  &$\displaystyle 0.82406  $ \\
&&&&\\
$c_{10}^{\text{int}}$ &&&&&& &$\displaystyle -\frac{0.22939 }{a}$ \\
&&&&\\
\hline
\hline
\end{tabular}

\end{table}

In this section, we consider  the Casimir energy inside a $D$-dimensional spherical cavity of radius $a$ with perfectly conducting boundary and with infinitely permeable boundary. As mentioned in the previous section, the Casimir energy has been computed in \cite{17} under the context of absolute and mixed boundary conditions using zeta regularization method.   Here we  compute the Casimir energy using cut-off regularization.

Using the cut-off regularization method, the Casimir energy inside a spherical shell of radius $a$ is given by
\begin{equation*}
E_{\text{Cas}}^{\text{int}} (a)=\frac{1}{2}\sum_{l=1}^{\infty}h_D(l)\sum_{j=1}^{\infty}\omega_{lj}^{\text{TE}}e^{-\lambda \omega_{lj}^{\text{TE}}}+
\frac{1}{2}\sum_{l=1}^{\infty}b_D(l)\sum_{j=1}^{\infty}\omega_{lj}^{\text{TM}}e^{-\lambda \omega_{lj}^{\text{TM}}},
\end{equation*}where $\lambda$ is a cut-off parameter.
Introducing the zeta function
\begin{equation}\label{eq3_3_5}
\begin{split}
\zeta^{\text{int}} (s)=\zeta_{\text{TE}}^{\text{int}} (s)+\zeta_{\text{TM}}^{\text{int}} (s)
=\sum_{l=1}^{\infty}h_D(l)\sum_{j=1}^{\infty}\left(\omega_{lj}^{\text{TE}}\right)^{-2s}+
 \sum_{l=1}^{\infty}b_D(l)\sum_{j=1}^{\infty}\left(\omega_{lj}^{\text{TM}}\right)^{-2s},
\end{split}
\end{equation}it is standard to show that up to  the $\lambda^0$ term,
\begin{align*}
E_{\text{Cas}}^{\text{int}} (a)=\sum_{i=1}^{D-1}\frac{\Gamma\left( D+1- i\right)}{\Gamma\left(\frac{D-i}{2}\right)}c_i^{\text{int}}\lambda^{i-D-1}-\frac{\psi(1)-\log\lambda}{2\sqrt{\pi}}c_{D+1}^{\text{int}}+\frac{1}{2} \text{FP}_{s=-\frac{1}{2}}\zeta^{\text{int}} (s),
\end{align*}where \begin{equation}\label{eq3_24_5}c_i^{\text{int}} =\text{Res}_{s=\frac{D-i}{2}}\left(\Gamma(s)\zeta^{\text{int}} (s)\right),\end{equation} and $\displaystyle \text{FP}_{s=-\frac{1}{2}}\zeta^{\text{int}} (s)$ is the finite part of the zeta function $\zeta^{\text{int}} (s)$ at $\displaystyle s=-\frac{1}{2}$. In the zeta regularization scheme, the terms with negative powers in $\lambda$ are omitted.
The zeta regularized Casimir energy is defined as
\begin{align*}
E_{\text{Cas}}^{  \text{int, zeta} }(a)=\left. \frac{\mu^{2s}}{2}\zeta^{\text{int}}\left(s-\frac{1}{2}\right)\right|_{s=0}=-\frac{c_{d+1} ^{\text{int}}}{4\sqrt{\pi}}\left(\frac{1}{\varepsilon}+\log\mu^2\right)+\frac{1}{2} \text{FP}_{s=-\frac{1}{2}}\zeta^{\text{int}} (s).
\end{align*}
Here $\mu$ is a normalization constant. The $1/\varepsilon+\log\mu^2$ term is the  ambiguity arises when $\zeta^{\text{int}} (s)$ has   pole at $\displaystyle s=-\frac{1}{2}$.
Systematic method  has been developed in \cite{30,31,15,17,32} for computing the  zeta functions   in spherical geometries. Since the computation is quite involved, we leave it to Appendix \ref{a1}.

\begin{table}[h]\caption{\label{t5}The value of $\displaystyle\text{FP}_{s=-1/2}\zeta^{\text{int}}(s)$  for the interior of a spherical shell of radius $a$ with perfectly conducting and infinitely permeable boundary conditions}

\begin{tabular}{c|c  c}
\hline
\hline
& perfectly conducting & infinitely permeable    \\
\hline
&&\\
$D=3$ & $\displaystyle\frac{0.16785}{a}+ \frac{16\log a}{315\pi a}$&$\displaystyle\frac{0.16785}{a}+ \frac{16\log a}{315\pi a}$\\
&&\\
$D=4$& $\displaystyle -\frac{0.04881}{a}-\frac{1631\log a}{65536a}$&$\displaystyle\frac{0.34619}{a}+ \frac{2713\log a}{65536 a}$\\
&&\\
$D=5$ & $\displaystyle \frac{0.01880}{a}+\frac{6632\log a}{45045\pi a}$ & $\displaystyle\frac{0.46773}{a} +\frac{11048\log a}{45045\pi a}$ \\
&&\\
$D=6$ & $\displaystyle -\frac{0.02022}{a}-\frac{1052991\log a}{116777216a}$ &  $\displaystyle\frac{0.52494}{a} +\frac{871339\log a}{8388608a}$\\
&&\\
$D=7$ & $\displaystyle\frac{0.01603}{a}+\frac{16438144\log a}{72747675\pi a} $ &  $\displaystyle\frac{0.56057}{a} +\frac{2955168\log a}{8083075\pi a}$\\
&&\\
$D=8$ & \hspace{0.5cm} $\displaystyle-\frac{0.00591}{a} -\frac{2580976217486942940567943799\log a}{32674585544991625157478973440 a}$ \hspace{0.5cm} & \hspace{0.5cm} $\displaystyle \frac{0.59822}{a} +\frac{4030878578159017023021286121\log a}{32674585544991625157478973440a}$ \hspace{0.5cm}\\
&&\\
$D=9$ & $\displaystyle-\frac{0.00358}{a}+\frac{0.08547\log a}{a}$ & $\displaystyle\frac{0.63741}{a}+\frac{0.12942\log a}{a}$\\
&&\\
\hline
\hline
\end{tabular}

\end{table}

Using the fact that the Hurwitz zeta function $\zeta_H(s;\chi)$ \eqref{eq3_24_4} has only one pole at $s=1$ with residue $1$, we can readily find the heat kernel coefficients
\eqref{eq3_24_5} from the expression for the zeta function $\zeta^{\text{int}}(s)$ \eqref{eq3_8_1}.  The results for $3\leq D\leq 9, 0\leq i\leq D+1$ are listed in Table \ref{t2} and Table \ref{t4}. The finite part of the zeta function $\zeta^{\text{int}}(s)$ at $\displaystyle s=-\frac{1}{2}$ can also be calculated directly from \eqref{eq3_8_1}. The result is listed in Table \ref{t5}.
  For $3\leq D\leq 6$, we find that our result is in good agreement with the result in \cite{17}. Notice that
  $\zeta^{\text{int}}(s)=a^{2s}\tilde{\zeta}^{\text{int}}(s),$ where  $\tilde{\zeta}^{\text{int}}(s)$ is independent of $a$. Therefore,
$$\text{FP}_{s=-1/2} \zeta^{\text{int}}(s) =\text{FP}_{s=-1/2} \tilde{\zeta}^{\text{int}}(s) +2\log a \text{Res}_{s=-1/2}\zeta^{\text{int}}(s)\\
=\text{FP}_{s=-1/2} \tilde{\zeta}^{\text{int}}(s) -\frac{c_{d+1}^{\text{int}}}{\sqrt{\pi}}\log a.$$The presence of the $\log a$ term indicates the ambiguity in defining zeta regularized Casimir energy.

\section{Casimir Energy in the Annular Region Between Spherical Shells}\label{s4}
In this section, we consider  the Casimir energy inside the  annular region between two spherical shells in $D$-dimensional space with radius $a$ and $b$, where $a<b$.  The Casimir energy  outside a $D$-dimensional spherical cavity with radius $a$ is obtained by taking the limit   $b\rightarrow \infty$.

As in the previous section, the Casimir energy inside the annular region between two spherical shells  with radius $a$ and $b$, where $a<b$, is given by
\begin{equation*}\begin{split}
E_{\text{Cas}}^{\text{ann}} (a,b)=&\frac{1}{2}\sum_{l=1}^{\infty}h_D(l)\sum_{j=1}^{\infty}\omega_{lj}^{\text{TE}}e^{-\lambda \omega_{lj}^{\text{TE}}}+
\frac{1}{2}\sum_{l=1}^{\infty}b_D(l)\sum_{j=1}^{\infty}\omega_{lj}^{\text{TM}}e^{-\lambda \omega_{lj}^{\text{TM}}}\\
=&\sum_{i=1}^{D-1}\frac{\Gamma\left( D+1- i\right)}{\Gamma\left(\frac{D-i}{2}\right)}c_i^{\text{ann}}\lambda^{i-D-1}-\frac{\psi(1)-\log\lambda}{2\sqrt{\pi}}c_{D+1}^{\text{ann}}+\frac{1}{2} \text{FP}_{s=-\frac{1}{2}}\zeta^{\text{ann}} (s).\end{split}
\end{equation*}As before, $$c_i^{\text{ann}} =\text{Res}_{s=\frac{D-i}{2}}\left(\Gamma(s)\zeta^{\text{ann}} (s)\right).$$
Using the same approach as in Appendix \ref{a1}, we find that for perfectly conducting boundary condition,
\begin{equation}\label{eq3_3_8}
\begin{split}
\zeta^{\text{ann}} (s)=\zeta_{\text{TE}}^{\text{ann}} (s)+\zeta_{\text{  TM}}^{\text{ann}} (s)
=\sum_{l=1}^{\infty}h_D(l)\zeta_D^{\text{ann}, l+\frac{D-2}{2}}(s)+
 \sum_{l=1}^{\infty}b_D(l)\zeta_{R,\frac{D-2}{2}}^{\text{ann},l+\frac{D-2}{2}}(s),
\end{split}
\end{equation} for infinitely permeable boundary condition,
\begin{equation}\label{eq3_3_9}
\begin{split}
\zeta^{\text{ann}} (s)=\zeta_{\text{TE}}^{\text{ann}} (s)+\zeta_{\text{TM}}^{\text{ann}} (s)
=\sum_{l=1}^{\infty}h_D(l)\zeta_{R,\frac{4-D}{2}}^{\text{ann}, l+\frac{D-2}{2}}(s)+
 \sum_{l=1}^{\infty}b_D(l)\zeta_{D}^{\text{ann}, l+\frac{D-2}{2}}(s),
\end{split}
\end{equation}where
\begin{equation*}
\begin{split}
\zeta_D^{\text{ann},  \nu}(s)=&\lim_{m\rightarrow 0}\frac{1}{2\pi i}\oint_{\gamma}(z^2+m^2)^{-s} \frac{d}{dz}\log\Bigl\{ J_{\nu}(az)N_{\nu}(bz)-J_{\nu}(bz)N_{\nu}(az)\Bigr\}dz,\\
\zeta_{R,c}^{ \text{ann}, \nu}(s)=&\lim_{m\rightarrow 0}\frac{1}{2\pi i}\oint_{\gamma}(z^2+m^2)^{-s}\\&\times \frac{d}{dz}\log\Bigl\{ \left[cJ_{\nu}(az)+azJ_{\nu}'(az)\right]
\left[cN_{\nu}(bz)+bzN_{\nu}'(bz)\right]-\left[cJ_{\nu}(bz)+bzJ_{\nu}'(bz)\right]\left[cN_{\nu}(az)+azN_{\nu}'(az)\right]\Bigr\}dz.
\end{split}
\end{equation*}   Using the fact that  $H_{\nu}^{(1)}(z)=J_{\nu}(z)+iN_{\nu}(z)$, $H_{\nu}^{(2)}(z)=J_{\nu}(z)-iN_{\nu}(z)$,  $\displaystyle J_{\nu}(iz)=e^{\frac{\pi \nu i}{2}}I_{\nu}(z), J_{\nu}(-iz)=e^{-\frac{\pi \nu i}{2}}I_{\nu}(z), H_{\nu}^{(1)}(iz)=\frac{2}{\pi i}e^{-\frac{\pi \nu i}{2}}K_{\nu}(z), H_{\nu}^{(2)}(-iz)=-\frac{2}{\pi i}e^{\frac{\pi \nu i}{2}}K_{\nu}(z)$, where $I_{\nu}(z)$ and $K_{\nu}(z)$ are the modified Bessel functions of the first kind and second kind  (see \cite{26}), one finds that
\begin{equation*}
\begin{split}
\zeta_D^{\text{ann},\nu}(s)=&\lim_{m\rightarrow 0}\frac{\sin (\pi s)}{\pi }\int_m^{\infty}(z^2-m^2)^{-s} \frac{d}{dz}\log\Bigl\{ I_{\nu}(bz)K_{\nu}(az)-I_{\nu}(az)K_{\nu}(bz)\Bigr\}dz,\\
\zeta_{R,c}^{\text{ann},\nu}(s)=&\lim_{m\rightarrow 0}\frac{\sin (\pi s)}{\pi }\int_m^{\infty}(z^2-m^2)^{-s}\\&\times \frac{d}{dz}\log\Bigl\{ \left[cI_{\nu}(az)+azI_{\nu}'(az)\right]\left[cK_{\nu}(bz)+bzK_{\nu}'(bz)\right]-\left[cI_{\nu}(bz)+bzI_{\nu}'(bz)\right]
\left[cK_{\nu}(az)+azK_{\nu}'(az)\right]\Bigr\}dz,
\end{split}
\end{equation*}for $\displaystyle \frac{1}{2}<\text{Re}\,s <1$. Notice that these can be written as
\begin{equation*}\begin{split}
\zeta_D^{\text{ann},\nu}(s)=\zeta_{D}^{\text{int},\nu}(s)+\zeta_{D}^{\text{ext},\nu}(s)+\zeta_{D}^{\text{inter},\nu}(s),\\
\zeta_{R,c}^{\text{ann}, \nu}(s)=\zeta_{R,c}^{\text{int},\nu}(s)+\zeta_{R,c}^{\text{ext},\nu}(s)+\zeta_{R,c}^{\text{inter},\nu}(s),
\end{split}\end{equation*}
where
\begin{equation*}\begin{split}
\zeta_{D}^{\text{int},\nu}(s)=&\lim_{m\rightarrow 0}\frac{\sin (\pi s)}{\pi }\int_m^{\infty}(z^2-m^2)^{-s} \frac{d}{dz}\log\Bigl\{ z^{-\nu}I_{\nu}(bz)\Bigr\}dz,\\
\zeta_{D}^{\text{ext},\nu}(s)=&\lim_{m\rightarrow 0}\frac{\sin (\pi s)}{\pi }\int_m^{\infty}(z^2-m^2)^{-s} \frac{d}{dz}\log\Bigl\{ z^{\nu}K_{\nu}(az)\Bigr\}dz,\\
\zeta_D^{\text{inter}, \nu}(s)=&\lim_{m\rightarrow 0}\frac{\sin (\pi s)}{\pi }\int_m^{\infty}(z^2-m^2)^{-s} \frac{d}{dz}\log\left\{ 1-\frac{I_{\nu}(az)K_{\nu}(bz)}{I_{\nu}(bz)K_{\nu}(az)}\right\}dz,\\
\zeta_{R,c}^{\text{int},\nu}(s)=&\lim_{m\rightarrow 0}\frac{\sin (\pi s)}{\pi }\int_m^{\infty}(z^2-m^2)^{-s} \frac{d}{dz}\log\Bigl\{ z^{-\nu}\left[cI_{\nu}(bz)+bzI_{\nu}'(bz)\right]\Bigr\}dz,\\
\zeta_{R,c}^{\text{ext},\nu}(s)=&\lim_{m\rightarrow 0}\frac{\sin (\pi s)}{\pi }\int_m^{\infty}(z^2-m^2)^{-s} \frac{d}{dz}\log\Bigl\{ z^{\nu}
\left[-cK_{\nu}(az)-azK_{\nu}'(az)\right]\Bigr\}dz,\\
\zeta_{R,c}^{\text{inter}, \nu}(s)=&\lim_{m\rightarrow 0}\frac{\sin (\pi s)}{\pi }\int_m^{\infty}(z^2-m^2)^{-s} \frac{d}{dz}\log\left\{  1-\frac{\left[cI_{\nu}(az)+azI_{\nu}'(az)\right]\left[cK_{\nu}(bz)+bzK_{\nu}'(bz)\right]}{\left[cI_{\nu}(bz)+bzI_{\nu}'(bz)\right]
\left[cK_{\nu}(az)+azK_{\nu}'(az)\right]} \right\}dz.
\end{split}\end{equation*}Correspondingly, the zeta function $\zeta^{\text{ann}}(s)$ \eqref{eq3_3_8} can be decomposed into three terms
$$\zeta^{\text{ann}}(s)=\zeta^{\text{int}}(s)+\zeta^{\text{ext}}(s)+\zeta^{\text{inter}}(s),$$which gives rise to
 the decomposition of the Casimir energy:
$$E_{\text{Cas}}^{\text{ann}} (a,b)=E_{\text{Cas}}^{\text{int}} (b)+E_{\text{Cas}}^{\text{ext}} (a)+E_{\text{Cas}}^{\text{inter}} (a,b).$$As seen from Appendix \ref{a1}, $E_{\text{Cas}}^{\text{int}}(b)$ is the Casimir energy inside a spherical shell of radius $b$. When $b\rightarrow \infty$, this can be considered as the Casimir energy of the free space. On the other hand,
  we show in Appendix \ref{a1} that in the limit $b\rightarrow \infty$,
\begin{align*}
\lim_{b\rightarrow \infty} \zeta_D^{\text{inter}, \nu}(s)=\lim_{b\rightarrow \infty} \zeta_{R,c}^{\text{inter}, \nu}(s)=0.
\end{align*}Therefore,   $E_{\text{Cas}}^{\text{ext}}(a)$ is  identified with the Casimir energy outside a spherical shell of radius $a$ (which has been renormalized by subtracting away the Casimir energy of the free space). We call $E_{\text{Cas}}^{\text{inter}} (a,b)$ the interacting term of the Casimir energy between the spherical shells.

 We consider the Casimir energy outside a spherical shell $ E_{\text{Cas}}^{\text{ext}}(a)$ and the interacting term $E_{\text{Cas}}^{\text{inter}} (a,b)$  separately in the following sections.

\section{Casimir Energy Outside a Spherical Shell }\label{s5}

The computation of the Casimir energy outside a spherical cavity of radius $a$ follows the same way as the computation for the Casimir energy inside the spherical cavity. We leave it to Appendix \ref{a1}. It can be shown explicitly that for $0\leq i\leq D-2$ and for $i=D-1, D+1$ and $D$ odd, the heat kernel coefficients for the exterior of the sphere $c_i^{\text{ext}}$ is related to the heat kernel coefficients for the interior of the sphere $c_i^{\text{int}}$ by
\begin{equation}\label{eq3_24_6}c_i^{\text{ext}}=(-1)^{i+1} c_i^{\text{int}}.\end{equation} For $i=D$ or $i=D-1, D+1$ and $D$ even, this is not necessary true. The values of $c_i^{\text{ext}}$ for $3\leq D\leq 9$ and $i=D-1, D, D+1,$ is listed in Table \ref{t6} and Table \ref{t7}. Compare to Table \ref{t2} and Table \ref{t4}, we find that \eqref{eq3_24_6} does not hold for $i=D$, but it  still holds for $i=D-1, D+1$ and $D$ even when $3\leq D\leq 9$. It seems that the latter will be true for all $D\geq 3$  but we don't know a proof for it when $D$ is even.

The value of the finite part of the zeta function $\zeta^{\text{ext}}(s)$ at $\displaystyle s=-\frac{1}{2}$ can be computed from \eqref{eq3_24_7} and the results for $3\leq D\leq 9$ are listed in Table \ref{t8}.

\begin{table} \caption{\label{t6}Heat kernel coefficients for the exterior of a spherical shell of radius $a$ with perfectly conducting boundary condition}

\begin{tabular}{c|ccccccc }
\hline
\hline
& $D=3$ & $D=4$ & $D=5$ & $D=6$ &  $D=7$ & $D=8$ & $D=9$  \\
\hline
&&&&&&\\
$c_{D-1}^{\text{int}}$ & $\displaystyle \frac{4a}{3\sqrt{\pi}}$ &$\displaystyle\frac{211\sqrt{\pi}a}{512}$ &$\displaystyle \frac{796a}{945\sqrt{\pi}}$&$\displaystyle \frac{75361\sqrt{\pi}a}{524288}$&$\displaystyle \frac{17468a}{61425\sqrt{\pi} }$&$ \displaystyle\frac{231850177\sqrt{\pi}a}{3019898880} $  &
$0.12483a$  \\
&&&&&&\\
$c_D^{\text{ext}}$ & $\displaystyle-\frac{3}{8}$ & $\displaystyle -\frac{13}{20}$ &$\displaystyle -\frac{2339}{2880}$ &$\displaystyle -\frac{6451}{7560}$ &$\displaystyle -\frac{824417}{967680}$ &$\displaystyle -\frac{193279}{226800}$ & $-0.85681$  \\
&&&&&&\\
$c_{D+1}^{\text{int}}$ & $\displaystyle \frac{16}{315\sqrt{\pi}a}$&$\displaystyle\frac{1631\sqrt{\pi}}{65536a}$&$\displaystyle \frac{6632}{45045\sqrt{\pi}a}$&$\displaystyle \frac{1052991\sqrt{\pi}}{16777216a}$&$\displaystyle\frac{16438144}{72747675\sqrt{\pi}a}$&$\displaystyle  \frac{2580976217486942940567943799\sqrt{\pi}}{32674585544991625157478973440a}$ & $\displaystyle\frac{0.15150}{a}$\\
&&&&\\
\hline
\hline
\end{tabular}

\end{table}

\begin{table} \caption{\label{t7}Heat kernel coefficients for the exterior of a spherical shell of radius $a$ with infinitely permeable boundary condition}

\begin{tabular}{c|ccccccc }
\hline
\hline
& $D=3$ & $D=4$ & $D=5$ & $D=6$ &  $D=7$ & $D=8$ & $D=9$  \\
\hline
&&&&&&\\
$c_{D-1}^{\text{int}}$ & $\displaystyle \frac{4a}{3\sqrt{\pi}}$ &$\displaystyle -\frac{121\sqrt{\pi}a}{512} $ &$\displaystyle \frac{68a}{189\sqrt{\pi}}$&$
\displaystyle- \frac{44071\sqrt{\pi}a}{524288}  $&$\displaystyle  \frac{58012a}{225225\sqrt{\pi}}$&$\displaystyle  - \frac{28775291\sqrt{\pi}a}{377487360}$  & $0.12063a$  \\
&&&&&&\\
$c_D^{\text{int}}$ & $\displaystyle-\frac{3}{8}$ & $\displaystyle \frac{11}{60} $ &$\displaystyle-\frac{451}{2880} $ &$\displaystyle \frac{1361}{7560}$ &$\displaystyle -\frac{182113}{967680} $ &$\displaystyle \frac{41351}{226800}$ & $-0.17594$ \\
&&&&&&\\
$c_{D+1}^{\text{int}}$ & $\displaystyle \frac{16}{315\sqrt{\pi}a}$&$\displaystyle -\frac{2713\sqrt{\pi}}{65536a}$&$\displaystyle \frac{11048}{45045\sqrt{\pi}a}$&$\displaystyle -\frac{871339\sqrt{\pi}}{8388608a}$&$\displaystyle  \frac{2955168}{8083075\sqrt{\pi}a} $&$\displaystyle-\frac{4030878578159017023021286121\sqrt{\pi}}{32674585544991625157478973440 a}$  & $\displaystyle\frac{0.22939}{a} $\\
&&&&\\
\hline
\hline
\end{tabular}

\end{table}

\begin{table} \caption{\label{t8}The value of $\displaystyle\text{FP}_{s=-1/2}\zeta^{\text{ext}}(s)$  for the exterior of a spherical shell of radius $a$ with perfectly conducting and infinitely permeable boundary conditions}

\begin{tabular}{c|c  c}
\hline
\hline
& perfectly conducting & infinitely permeable    \\
\hline
&&\\
$D=3$ & $\displaystyle-\frac{0.07549}{a}- \frac{16\log a}{315\pi a}$&$\displaystyle-\frac{0.07549}{a}- \frac{16\log a}{315\pi a}$\\
&&\\
$D=4$& $\displaystyle -\frac{0.18230}{a}-\frac{1631\log a}{65536}$&$\displaystyle\frac{0.02446}{a}+ \frac{2713\log a}{65536 a}$\\
&&\\
$D=5$ & $\displaystyle -\frac{0.29414}{a}-\frac{6632\log a}{45045\pi a}$ & $\displaystyle-\frac{0.01042}{a} -\frac{11048\log a}{45045\pi a}$ \\
&&\\
$D=6$ & $\displaystyle -\frac{0.38128}{a}-\frac{1052991\log a}{116777216a}$ &  $\displaystyle\frac{0.00447}{a} +\frac{871339\log a}{8388608a}$\\
&&\\
$D=7$ & $\displaystyle-\frac{0.44667}{a}-\frac{16438144\log a}{72747675\pi a} $ &  $\displaystyle\frac{0.00566}{a} -\frac{2955168\log a}{8083075\pi a}$\\
&&\\
$D=8$ & \hspace{0.5cm} $\displaystyle-\frac{0.50087}{a} -\frac{2580976217486942940567943799\log a}{32674585544991625157478973440 a}$ \hspace{0.5cm} & \hspace{0.5cm} $\displaystyle -\frac{0.01784}{a} +\frac{4030878578159017023021286121\log a}{32674585544991625157478973440a}$ \hspace{0.5cm}\\
&&\\
$D=9$ & $\displaystyle -\frac{0.54880}{a}-\frac{0.08547\log a}{a}$ & $\displaystyle\frac{0.02920}{a}-\frac{0.12942\log a}{a}$\\
&&\\
\hline
\hline
\end{tabular}

\end{table}

\section{  Total Casimir Energy of a Spherical Shell}\label{s6}
\begin{table} \caption{\label{t5_3}The value of $\displaystyle\frac{1}{2}\text{FP}_{s=-1/2}\zeta^{\text{single}}(s)$ for a spherical shell of radius $a$ with perfectly conducting and infinitely permeable boundary conditions}

\begin{tabular}{c|c  c}
\hline
\hline
& perfectly conducting & infinitely permeable    \\
\hline
&&\\
$D=3$ & $\displaystyle\frac{0.04618}{a} $&$\displaystyle\frac{0.04618}{a} $\\
&&\\
$D=4$& $\displaystyle -\frac{0.11555}{a}-\frac{1631\log a}{65536}$&$\displaystyle\frac{0.18533}{a}+ \frac{2713\log a}{65536 a}$\\
&&\\
$D=5$ & $\displaystyle -\frac{0.13767}{a} $ & $\displaystyle\frac{0.22866}{a}  $ \\
&&\\
$D=6$ & $\displaystyle -\frac{0.20075}{a}-\frac{1052991\log a}{116777216a}$ &  $\displaystyle\frac{0.26470}{a} +\frac{871339\log a}{8388608a}$\\
&&\\
$D=7$ & $\displaystyle-\frac{0.21532}{a} $ &  $\displaystyle\frac{0.28312}{a}  $\\
&&\\
$D=8$ & \hspace{0.5cm} $\displaystyle-\frac{0.25339}{a} -\frac{2580976217486942940567943799\log a}{32674585544991625157478973440 a}$ \hspace{0.5cm} & \hspace{0.5cm} $\displaystyle \frac{0.29019}{a} +\frac{4030878578159017023021286121\log a}{32674585544991625157478973440a}$ \hspace{0.5cm}\\
&&\\
$D=9$ & $\displaystyle -\frac{0.27619}{a}$ & $\displaystyle \frac{0.33331}{a}$\\
&&\\
\hline
\hline
\end{tabular}

\end{table}

In this section, we consider the Casimir energy of a single spherical shell system.
The Casimir energy for  a perfectly conducting spherical shell or an infinitely permeable spherical shell is the sum of the Casimir energy inside the spherical shell and the Casimir energy outside the spherical shell. In the cut-off scheme, we find that
\begin{align*}
E_{\text{Cas}}^{\text{single}} (a)=\sum_{i=1}^{D-1}\frac{\Gamma\left( D+1- i\right)}{\Gamma\left(\frac{D-i}{2}\right)}c_i^{\text{single}} \lambda^{i-D-1}-\frac{\psi(1)-\log\lambda}{2\sqrt{\pi}}c_{D+1}^{\text{single}} +\frac{1}{2} \text{FP}_{s=-\frac{1}{2}}\zeta^{\text{single}}  (s),
\end{align*}up to the term constant in $\lambda$. Here
\begin{equation*}
\zeta^{\text{single}}(s)=\zeta^{\text{int}}(s)+\zeta^{\text{ext}}(s), \hspace{1cm} c_i^{\text{single}}=c_i^{\text{int}}+c_i^{\text{ext}}.
\end{equation*}\eqref{eq3_24_6} shows that if $i$ is even and $i\neq D$, $c_i^{\text{single}}=0$, but  if $i$ is odd, $c_i^{\text{single}} $  is in general nonzero. However when $D=3$, $c_1^{\text{single}}=0$ since both $c_1^{\text{int}}$ and $c_1^{\text{ext}}$ vanish. Therefore, when $D=3$, $c_0^{\text{single}}, c_1^{\text{single}}, c_2^{\text{single}}, c_4^{\text{single}}$ all vanish and one can take $\lambda\rightarrow 0^+$ in the cut off regularized Casimir energy to obtain an unambiguous Casimir energy given by $\displaystyle\frac{1}{2}\text{FP}_{s=-1/2}\zeta^{\text{single}}(s)$ whose numerical value is well-known to be $\displaystyle\frac{0.04618}{a} $ correct to five decimal places.  When $D\geq 4$,   divergence is  always present in the cut-off regularized Casimir energy.  The leading divergence is  of order $\lambda^{-D}$. In order to obtain a physically meaningful vacuum energy, one need to remove the divergence by some renormalization procedures.

In the zeta regularized scheme, the Casimir energy is defined as
\begin{align*}
E_{\text{Cas}}^{  \text{single,  zeta} }(a)=\left. \frac{\mu^{2s}}{2}\zeta^{\text{single}} \left(s-\frac{1}{2}\right)\right|_{s=0}=-\frac{c_{d+1}^{\text{single}} }{4\sqrt{\pi}}\left(\frac{1}{\varepsilon}+\log\mu^2\right)+\frac{1}{2} \text{FP}_{s=-\frac{1}{2}}\zeta^{\text{single}}  (s).
\end{align*}
The value of $\displaystyle\frac{1}{2}\text{FP}_{s=-1/2}\zeta^{\text{single}}(s)$ for $3\leq D\leq 9$ is tabulated in Table \ref{t5_3}. As mentioned in the previous section, when $D$ is odd, it can be shown that $c_{D+1}^{\text{ext}}=-c_{D+1}^{\text{int}}$. Therefore, $c_{D+1}^{\text{single}}=0$ when $D$ is odd. In this case, the zeta regularized Casimir energy is finite and  given by $\displaystyle\frac{1}{2}\text{FP}_{s=-1/2}\zeta^{\text{single}}(s)$.

The Casimir force acting on a single spherical shell is given by
\begin{align*}
F_{\text{Cas}}^{  \text{single } }(a)=-\frac{\pa}{\pa a} E_{\text{Cas}}^{  \text{   single} }(a).
\end{align*}Using cut-off regularization, we find that there are divergences in the Casimir force for $D\geq 4$. Using zeta regularization scheme, there is no divergence when $D$ is odd, and the Casimir force is given by
\begin{align*}
F_{\text{Cas}}^{  \text{single, zeta } }(a)= \frac{E_{\text{Cas}}^{  \text{   single, zeta}  }(a)}{a}.
\end{align*}When $D=3$, we have the well-known result that the Casimir force is repulsive. However, for $4\leq D\leq 9$ an odd dimension, we notice from Table
\ref{t5_3} that the Casimir force is attractive for perfectly conducting boundary condition, and repulsive for infinite permeable boundary condition.

\section{  Casimir Effect of Double Spherical Shell Configuration}\label{s7}
In this section, we consider two concentric spherical shells of radius $a$ and $b$ respectively, where $a<b$. The total Casimir energy of such a system is the sum of the Casimir energy inside the sphere of radius $a$, the Casimir energy inside the annular region bounded by the two spheres, and the Casimir energy outside the sphere of radius $b$. Namely,
\begin{equation*}
E_{\text{Cas}}^{\text{double}}(a,b)=E_{\text{Cas}}^{\text{int}}(a )+E_{\text{Cas}}^{\text{ann}}(a,b)+E_{\text{Cas}}^{\text{ext}}( b).
\end{equation*}As   discussed in Section \ref{s4}, the Casimir energy in the annular region $E_{\text{Cas}}^{\text{ann}}(a,b)$ can be decomposed into
\begin{equation*}
E_{\text{Cas}}^{\text{ann}}(a,b)=E_{\text{Cas}}^{\text{int}} (b)+E_{\text{Cas}}^{\text{ext}} (a)+E_{\text{Cas}}^{\text{inter}} (a,b).
\end{equation*}Therefore, we can rewrite the total Casimir energy of a double spherical shell system as
\begin{equation*}
E_{\text{Cas}}^{\text{double}}(a,b)=E_{\text{Cas}}^{\text{single}}(a )+E_{\text{Cas}}^{\text{inter}}(a,b)+E_{\text{Cas}}^{\text{single}}( b),
\end{equation*}which is the sum of the Casimir energy of a single spherical shell  of radius $a$, the Casimir energy of a single spherical shell  of radius $b$, and the interacting term $E_{\text{Cas}}^{\text{inter}}(a,b)$.
The Casimir energy of a single spherical shell  has been discussed in the previous section. We find that using cut-off regularization, it does not have divergence only when $D=3$. Using zeta regularization, the Casimir energy does not have divergence for all odd $D$. When $D=3$, both the cut-off regularization and zeta regularization yield the same result.

Using cut-off regularization, we have as before,
\begin{align*}
E_{\text{Cas}}^{\text{double}} (a)=\sum_{i=1}^{D-1}\frac{\Gamma\left( D+1- i\right)}{\Gamma\left(\frac{D-i}{2}\right)}c_i^{\text{double}} \lambda^{i-D-1}-\frac{\psi(1)-\log\lambda}{2\sqrt{\pi}}c_{D+1}^{\text{double}} +\frac{1}{2} \text{FP}_{s=-\frac{1}{2}}\zeta^{\text{double}}  (s),
\end{align*}up to the term constant in $\lambda$. Here
$$ \zeta^{\text{double}}(s)=\zeta^{\text{single}}(s;a)+\zeta^{\text{single}}(s;b)+\zeta^{\text{inter}}(s).$$
In Appendix \ref{a1}, we show   that $\zeta^{\text{inter}}(s)$ is analytic for all $s$. Therefore, when $0\leq i\leq D-1$ or $i=D+1$, the contribution to
$$c_i^{\text{double}}=\text{Res}_{s=\frac{D-i}{2}}\left(\Gamma(s)\zeta^{\text{double}}(s)\right)$$ comes only from $\zeta^{\text{single}}(s;a)$ and $\zeta^{\text{single}}(s;b)$. Therefore, the $\lambda\rightarrow 0^+$ divergence term of the Casimir energy $E_{\text{Cas}}^{\text{double}} (a)$ is a sum of the divergence term for a single spherical shell of radius $a$ and the divergence term for a single spherical shell of radius $b$. As a result, the divergence vanishes only if  $D=3$.

Since the single shell contributions to the Casimir energy have been discussed in the previous section,    we focus our attention on the interacting term. From the result of \eqref{eq3_25_2} in Appendix \ref{a1},   we find that for perfectly conducting boundary condition, the interacting term of the Casimir energy is
\begin{equation} \label{eq3_25_3}
\begin{split}
E_{\text{Cas}}^{\text{inter}}(a,b)=&\frac{1}{2}\zeta^{\text{inter}}\left(-\frac{1}{2}\right)=\frac{1}{2\pi}\sum_{l=1}^{\infty}h_D(l)\int_0^{\infty}\log\left\{ 1-\frac{I_{\nu(l)}(az)K_{\nu(l)}(bz)}{I_{\nu(l)}(bz)K_{\nu(l)}(az)}\right\}dz
\\&+\frac{1}{2\pi}\sum_{l=1}^{\infty}b_D(l)\int_0^{\infty}\log\left\{  1-\frac{\left[\tfrac{D-2}{2}I_{\nu(l)}(az)+azI_{\nu(l)}'(az)\right]\left[\tfrac{D-2}{2}K_{\nu(l)}(bz)+bzK_{\nu(l)}'(bz)\right]}{\left[\tfrac{D-2}{2}I_{\nu(l)}(bz)+bzI_{\nu(l)}'(bz)\right]
\left[\tfrac{D-2}{2}K_{\nu(l)}(az)+azK_{\nu(l)}'(az)\right]} \right\}dz.
\end{split}
\end{equation} For infinitely permeable boundary condition,
\begin{equation}\label{eq3_25_4}
\begin{split}
E_{\text{Cas}}^{\text{inter}}(a,b)=&\frac{1}{2\pi}\sum_{l=1}^{\infty}h_D(l)\int_0^{\infty}\log\left\{  1-\frac{\left[\tfrac{4-D}{2}I_{\nu(l)}(az)+azI_{\nu(l)}'(az)\right]\left[\tfrac{4-D}{2}K_{\nu(l)}(bz)+bzK_{\nu(l)}'(bz)\right]}{\left[
\tfrac{4-D}{2}I_{\nu(l)}(bz)+bzI_{\nu(l)}'(bz)\right]
\left[\tfrac{4-D}{2}K_{\nu(l)}(az)+azK_{\nu(l)}'(az)\right]} \right\}dz
\\&+\frac{1}{2\pi}\sum_{l=1}^{\infty}b_D(l)\int_0^{\infty}\log\left\{ 1-\frac{I_{\nu(l)}(az)K_{\nu(l)}(bz)}{I_{\nu(l)}(bz)K_{\nu(l)}(az)}\right\}dz.
\end{split}
\end{equation}
  The contribution of the interacting term to the Casimir force acting on the shell with radius $a$ is given by
$$F_{\text{Cas}}^{a,\text{inter}}(a,b)=-\frac{\pa}{\pa a} E_{\text{Cas}}^{\text{inter}}(a,b).$$ In case of perfectly conducting boundary condition,
\begin{equation*}
\begin{split}
F_{\text{Cas}}^{a,\text{inter}}(a,b)=&\frac{1}{2\pi a}\sum_{l=1}^{\infty}h_l(D)\int_0^{\infty}\Delta_{D}^{a,l+\frac{D-2}{2} }(z;a,b)dz+\frac{1}{2\pi a}\sum_{l=1}^{\infty} b_l(D)
\int_0^{\infty}\Delta_{R,\frac{D-2}{2}}^{a,l+\frac{D-2}{2}}(z;a,b)dz.
\end{split}
\end{equation*}In case of infinitely permeable boundary condition,
\begin{equation*}
\begin{split}
F_{\text{Cas}}^{a,\text{inter}}(a,b)=&\frac{1}{2\pi a}\sum_{l=1}^{\infty}h_l(D)\int_0^{\infty}\Delta_{R,\frac{4-D}{2}}^{a,l+\frac{D-2}{2} }(z;a,b)dz+\frac{1}{2\pi a}\sum_{l=1}^{\infty} b_l(D)
\int_0^{\infty}\Delta_{D}^{a,l+\frac{D-2}{2}}(z;a,b)dz.
\end{split}
\end{equation*}Here
\begin{equation*}\begin{split}
\Delta_{D}^{a,\nu(l) }(z;a,b)=& \left.\frac{K_{\nu(l)}(bz)}{I_{\nu(l)}(bz)K_{\nu(l)}(az)^2}\right/\left(1-\frac{I_{\nu(l)}(az)K_{\nu(l)}(bz)}{I_{\nu(l)}(bz)K_{\nu(l)}(az)}\right). \\
\Delta_{R,c}^{a,\nu(l)}(z;a,b)=&\left.  \tfrac{ -\left[cK_{\nu(l)}(bz)+bzK_{\nu(l)}'(bz)\right]\left(a^2z^2+(\nu^2-c^2)\right)}
{\left[cI_{\nu(l)}(bz)+bzI_{\nu(l)}'(bz)\right]\left[cK_{\nu(l)}(az)+azK_{\nu(l)}'(az)\right]^2}\right/
\left(1-\tfrac{\left[cI_{\nu(l)}(az)+azI_{\nu(l)}'(az)\right]\left[cK_{\nu(l)}(bz)+bzK_{\nu(l)}'(bz)\right]}
{\left[cI_{\nu(l)}(bz)+bzI_{\nu(l)}'(bz)\right]\left[cK_{\nu(l)}(az)+azK_{\nu(l)}'(az)\right]}\right).
\end{split}
\end{equation*}Notice that the function $$cI_{\nu(l)}(z)+zI_{\nu(l)}'(z)=(c+\nu(l))I_{\nu(l)}(z)+zI_{\nu(l)+1}(z)$$ is $\geq 0$ when $z\geq 0$. For $\displaystyle c=\tfrac{D-2}{2}$ or $\displaystyle \tfrac{4-D}{2}$ and $\displaystyle \nu(l)=l+\tfrac{D-2}{2}$, $\nu(l)-c$ is always positive. Therefore the function
$$cK_{\nu(l)}(z)+zK_{\nu(l)}'(z)= (c-\nu(l))K_{\nu(l)}(z)-zK_{\nu(l)-1}(z)$$ is $\leq 0$ when $z\geq 0$. On the other hand, since
\begin{equation*}
\begin{split}
\frac{d}{dz}\frac{K_{\nu(l)}(z)}{I_{\nu(l)}(z)}=&-\frac{1}{zI_{\nu(l)}(z)^2}\leq 0,\\
\frac{d}{dz}\left( -\frac{cK_{\nu(l)}( z)+ zK_{\nu(l)}'( z)}{cI_{\nu(l)}( z)+ zI_{\nu(l)}'( z)}\right)=&-\frac{z^2+(\nu(l)^2-c^2)}{z\left(cI_{\nu(l)}( z)+ zI_{\nu(l)}'( z)
\right)^2}\leq 0,
\end{split}
\end{equation*}  $\displaystyle  K_{\nu(l)}(z)/I_{\nu(l)}(z) $ and $\displaystyle  -\left.\left(cK_{\nu(l)}( z)+ zK_{\nu(l)}'( z)\right)\right/\left(cI_{\nu(l)}( z)+ zI_{\nu(l)}'( z)\right)$ are nonnegative decreasing functions of $z$ when $z\geq 0$. Therefore, when $z\geq 0$,     $\displaystyle \nu(l)=l+\tfrac{D-2}{2}$ and $\displaystyle c=\tfrac{D-2}{2}$ or $\displaystyle \tfrac{4-D}{2}$, we find that
\begin{equation}\label{eq3_25_8}
0\leq \frac{ K_{\nu(l)}(bz)}{I_{\nu(l)}(bz) }\leq \frac{ K_{\nu(l)}(az)}{I_{\nu(l)}(az) },
\hspace{1cm}
0\leq -\frac{ cK_{\nu(l)}(bz)+bzK_{\nu(l)}'(bz) }
{ cI_{\nu(l)}(bz)+bzI_{\nu(l)}'(bz) }\leq -\frac{ cK_{\nu(l)}(az)+azK_{\nu(l)}'(az) }
{ cI_{\nu(l)}(az)+azI_{\nu(l)}'(az) }.
\end{equation}
 These imply that
 $\Delta_{D}^{a,\nu(l) }(z;a,b)$ and $\Delta_{R,c}^{a,\nu(l)}(z;a,b)$ are nonnegative functions of $z$ when $z\geq 0$. As a result, we find that for either perfectly conducting or infinitely permeable boundary conditions, the interacting term of the Casimir energy gives rise to a force acting on the inner shell that tend to push the inner shell towards the outer shell.

Next we consider the interacting term of the Casimir force acting on the outer shell with radius $b$. This is given by
\begin{align*}
F_{\text{Cas}}^{b,\text{inter}}(a,b)=-\frac{\pa}{\pa b} E_{\text{Cas}}^{\text{inter}}(a,b).
\end{align*}For  perfectly conducting boundary condition and infinitely permeable boundary condition, we have  respectively
\begin{equation}\label{eq3_31_1}
\begin{split}
F_{\text{Cas}}^{b,\text{inter}}(a,b)=&-\frac{1}{2\pi b}\sum_{l=1}^{\infty}h_l(D)\int_0^{\infty}\Delta_{D}^{b,l+\frac{D-2}{2} }(z;a,b)dz-\frac{1}{2\pi b}\sum_{l=1}^{\infty} b_l(D)
\int_0^{\infty}\Delta_{R,\frac{D-2}{2}}^{b,l+\frac{D-2}{2}}(z;a,b)dz.
\end{split}
\end{equation}and
\begin{equation}\label{eq3_31_2}
\begin{split}
F_{\text{Cas}}^{b,\text{inter}}(a,b)=&-\frac{1}{2\pi b}\sum_{l=1}^{\infty}h_l(D)\int_0^{\infty}\Delta_{R,\frac{4-D}{2}}^{b,l+\frac{D-2}{2} }(z;a,b)dz-\frac{1}{2\pi b}\sum_{l=1}^{\infty} b_l(D)
\int_0^{\infty}\Delta_{D}^{b,l+\frac{D-2}{2}}(z;a,b)dz.
\end{split}
\end{equation}
Here
\begin{equation*}\begin{split}
\Delta_{D}^{b,\nu(l) }(z;a,b)=& \left.\frac{I_{\nu(l)}(az)}{K_{\nu(l)}(az)I_{\nu(l)}(bz)^2}\right/\left(1-\frac{I_{\nu(l)}(az)K_{\nu(l)}(bz)}{I_{\nu(l)}(bz)K_{\nu(l)}(az)}\right). \\
\Delta_{R,c}^{b,\nu(l)}(z;a,b)=&\left.  \tfrac{  \left[cI_{\nu(l)}(az)+azI_{\nu(l)}'(az)\right]\left(b^2z^2+(\nu^2-c^2)\right)}
{-\left[cK_{\nu(l)}(az)+azK_{\nu(l)}'(az)\right]\left[cI_{\nu(l)}(bz)+bzI_{\nu(l)}'(bz)\right]^2}\right/
\left(1-\tfrac{\left[cI_{\nu(l)}(az)+azI_{\nu(l)}'(az)\right]\left[cK_{\nu(l)}(bz)+bzK_{\nu(l)}'(bz)\right]}
{\left[cI_{\nu(l)}(bz)+bzI_{\nu(l)}'(bz)\right]\left[cK_{\nu(l)}(az)+azK_{\nu(l)}'(az)\right]}\right).
\end{split}
\end{equation*}It follows from \eqref{eq3_25_8} that $\Delta_{D}^{b,\nu(l) }(z;a,b)$ and $\Delta_{R,c}^{a,\nu(l)}(z;a,b)$ are always nonnegative when $z\geq 0$,     $\displaystyle \nu(l)=l+\tfrac{D-2}{2}$ and $\displaystyle c=\tfrac{D-2}{2}$ or $\displaystyle \tfrac{4-D}{2}$. Therefore, the minus signs in front of the integrals in \eqref{eq3_31_1} and \eqref{eq3_31_2} imply that the interacting term of the Casimir energy always gives rise to a force acting on the outer shell that tends to push the outer shell towards the inner shell. Combining with the result about the interacting term of the Casimir force on the inner shell, we find that the interacting term of the Casimir energy gives rise to a Casimir force that attracts the two shells to each other.
Taking into account the single shell contribution, the Casimir force can only be unambiguously regularized in the context of cut-off regularization when the space dimension is $D=3$. In this case, we see that   acting on the inner shell with radius $a$, the total Casimir force is
$$F_{\text{Cas}}^a (a,b)=F^{\text{single}}_{\text{Cas}}(a)+F_{\text{Cas}}^{a,\text{inter}}(a,b)=\frac{0.04618}{a^2}+F_{\text{Cas}}^{a,\text{inter}}(a,b),$$ Therefore it is always positive (pointing outward). However, for the outer shell of radius $b$, the total Casimir force is
$$F_{\text{Cas}}^b (a,b)=F^{\text{single}}_{\text{Cas}}(b)+F_{\text{Cas}}^{b,\text{inter}}(a,b)=\frac{0.04618}{a^2}+F_{\text{Cas}}^{b,\text{inter}}(a,b).$$The first term is positive but the second term is negative. It is easy to show that when $b\rightarrow a^+$, $\Delta_{D}^{b,\nu(l) }(z;a,b)$ and $\Delta_{R,c}^{b,\nu(l)}(z;a,b)$ approach positive infinity, and when $b\rightarrow \infty$, they approach zero exponentially fast. Therefore, for $b$ close to $a$,  the Casimir force acting on the outer shell  $F_{\text{Cas}}^b (a,b)$ is negative (pointing inward). However, for fixed $a$, when $b$ is large enough, the Casimir force will become positive. There is an equilibrium point where the force is zero but it is unstable. Numerical simulation shows that the equilibrium point appears at $b/a=5.55857$ correct to five decimal places.
If the space dimension $D$ is odd, one can also discuss the sign of the total Casimir force acting on the two shells in the similar way under the context of zeta regularization.

\section{Asymptotic Behavior of the  Casimir Force for Small Separation of Shells }\label{s8}
In this section, we consider the asymptotic behavior of the  Casimir force when the separation between the plates $d=b-a$ is small compared to the radii of the spheres. Since the single shell contribution to the Casimir force does not depend on the plate separation, we only consider the interacting term.  We first consider the force acting on the inner shell.
By a change of variable, we find that in case of perfectly conducting boundary conditions, the interacting term is given by
\begin{equation}\label{eq3_26_3}
\begin{split}
F_{\text{Cas}}^{a,\text{inter}}(a,b)=&\frac{1}{2\pi a^2}\sum_{l=1}^{\infty}\nu(l) h_l(D)\int_0^{\infty}\Delta_{D}^{a,l+\frac{D-2}{2} }\left(\tfrac{\nu (l) z}{a};1,\tfrac{b}{a}\right)dz+\frac{1}{2\pi a^2}\sum_{l=1}^{\infty}\nu(l) b_l(D)
\int_0^{\infty}\Delta_{R,\frac{D-2}{2}}^{a,l+\frac{D-2}{2}}\left(\tfrac{\nu (l) z}{a};1,\tfrac{b}{a}\right)dz.
\end{split}
\end{equation} Using Debye uniform asymptotic expansions of the Bessel functions \eqref{eq3_4_1} and \eqref{eq3_15_2}, we find that as $\nu\rightarrow \infty$,
\begin{equation}\label{eq3_26_4}
\begin{split}
\Delta_{D}^{a,\nu }\left(\tfrac{\nu z}{a};1,\tfrac{b}{a}\right)
\sim & 2\nu \sqrt{1+z^2}\sum_{n=1}^{\infty}e^{-2n\nu \left(\eta\left(\frac{bz}{a}\right)-\eta(z)\right)}\left(1+\sum_{i=1}^{\infty}\frac{p_{ni}(z;a,b)}{\nu^i}\right),\\
\Delta_{R,\frac{D-2}{2}}^{a,\nu}\left(\tfrac{\nu  z}{a};1,\tfrac{b}{a}\right)\sim & 2 \frac{\nu^2z^2+(\nu^2-c^2)}{\nu \sqrt{1+z^2}}\sum_{n=1}^{\infty}e^{-2n\nu \left(\eta\left(\frac{bz}{a}\right)-\eta(z)\right)}\left(1+\sum_{i=1}^{\infty}\frac{q_{ni}(z;a,b)}{\nu^i}\right).
\end{split}
\end{equation}$p_{ni}(z;a,b)$ and $q_{ni}(z;a,b)$ are polynomials of $t(z)=1/\sqrt{1+z^2}$ and $t\left(bz/a\right)$ that vanish when $b\rightarrow a$. Using the convention $p_{n0}(z;a,b)=q_{n0}(z;a,b)\equiv 1$,
 \eqref{eq3_26_4} implies that
\begin{equation}\label{eq3_26_5}
\begin{split}
&\frac{1}{2\pi a^2}\sum_{l=1}^{\infty}\nu(l) h_l(D)\int_0^{\infty}\Delta_{D}^{a,l+\frac{D-2}{2} }\left(\tfrac{\nu (l) z}{a};1,\tfrac{b}{a}\right)dz
\\\sim & \frac{1}{ \pi a^2}\sum_{l=1}^{\infty}\nu(l)^2 h_l(D)\int_0^{\infty}\sqrt{1+z^2}\sum_{n=1}^{\infty}e^{-2n\nu(l) \left(\eta\left(\frac{bz}{a}\right)-\eta(z)\right)} \sum_{i=0}^{\infty}\frac{p_{ni}(z;a,b)}{\nu^i} dz\\
=&\frac{1}{ \pi a^2}\sum_{j=0}^{D-2}x_{D;j}\sum_{l=1}^{\infty}\nu(l)^{j+2}  \int_0^{\infty}\sqrt{1+z^2}
\frac{1}{2\pi i}\int_{\mu-i\infty}^{\mu +i\infty} \Gamma(s)(2n\nu(l))^{-s}  \left(\eta\left(\frac{bz}{a}\right)-\eta(z)\right)^{-s}ds\sum_{i=0}^{\infty}\frac{p_{ni}(z;a,b)}{\nu^i} dz\\
=&\frac{1}{\pi a^2}\frac{1}{2\pi i} \int_{\mu-i\infty}^{\mu +i\infty} \Gamma(s)2^{-s}\sum_{i=0}^{\infty}\sum_{n=1}^{\infty} n^{-s}\sum_{j=0}^{D-2}x_{D;j}\zeta_H\left(s+i-j-2;\tfrac{D}{2}\right)\int_0^{\infty}\sqrt{1+z^2}\left(\eta\left(\frac{bz}{a}\right)-\eta(z)\right)^{-s}p_{ni}(z;a,b)dz ds
\end{split}
\end{equation}for large enough $\mu$.
Notice that
\begin{equation*}
\begin{split}
\eta\left(\frac{bz}{a}\right)-\eta(z) &\rightarrow \log\frac{b}{a}\hspace{1cm}\text{as}\;\;z\rightarrow 0\\
\eta\left(\frac{bz}{a}\right)-\eta(z) &\sim \left(\frac{b}{a}-1\right)z\hspace{1cm}\text{as}\;\;z\rightarrow \infty.
\end{split}
\end{equation*}Therefore the integral in $z$ in the last line of  \eqref{eq3_26_5} is convergent as long as $s>2$.
Let
$$\gamma=\frac{b}{a}-1=\frac{d}{a}.$$  The leading term of \eqref{eq3_26_5} when $\gamma\rightarrow 0$ is given by the term with $i=0, j=D-2$ which gives rise to a pole of $\zeta_H\left(s+i-j-2;\tfrac{D}{2}\right)$ at $s=D+1$. We find that this leading term is
\begin{equation}\label{eq3_26_6}
\begin{split}
 \frac{1}{\pi a^2}\frac{\Gamma(D+1)}{2^{D+1}}\sum_{n=1}^{\infty} n^{-D-1}x_{D;D-2}\int_0^{\infty} \sqrt{1+z^2}\left(\eta\left(\frac{bz}{a}\right)-\eta(z)\right)^{-D-1}dz.
\end{split}
\end{equation}By definition $x_{D;j}=2/(D-3)!$. On the other hand, to the first order in $\gamma$, $$ \eta\left(\frac{bz}{a}\right)-\eta(z)=\left(\frac{bz}{a}-z\right)
\eta'(z)+O\left(\gamma^2\right)=\gamma\sqrt{1+z^2} +O\left(\gamma^2\right).$$ Therefore, the leading term of \eqref{eq3_26_6} when $\gamma\rightarrow 0$ is given by
\begin{equation*}
\begin{split}
 \frac{1}{\pi a^2\gamma^{D+1}}\frac{\Gamma(D+1)}{2^{D+1}}\zeta_R(D+1)\frac{2}{(D-3)!} \int_0^{\infty}\frac{1}{(1+z^2)^{\frac{D}{2}}}dz= \frac{1}{\sqrt{\pi} a^2\gamma^{D+1}}\frac{\Gamma(D+1)}{2^{D+1}}\frac{\zeta_R(D+1)}{(D-3)!}\frac{\Gamma\left(\frac{D-1}{2}\right)}{\Gamma\left(\frac{D}{2}\right)}.
\end{split}
\end{equation*}In the similar way, one can show that the leading contribution of the second term in \eqref{eq3_26_3} when $\gamma\rightarrow 0$ is given by
$$ \frac{1}{\sqrt{\pi} a^2\gamma^{D+1}}\frac{\Gamma(D+1)}{2^{D+1}}\frac{\zeta_R(D+1)}{(D-2)!}\frac{\Gamma\left(\frac{D-1}{2}\right)}{\Gamma\left(\frac{D}{2}\right)}.$$
Summing up, we find that as the separation between the shells $d$ goes to zero, the leading term of the Casimir force is
\begin{equation}\label{eq3_26_7}\begin{split}
F_{\text{Cas}}^{a,\text{inter}}(a,b)\sim &\frac{1}{\sqrt{\pi} a^2\gamma^{D+1}}\frac{\Gamma(D+1)}{2^{D+1}}\frac{\zeta_R(D+1)}{(D-3)!}\frac{\Gamma\left(\frac{D-1}{2}\right)}{\Gamma\left(\frac{D}{2}\right)}
+\frac{1}{\sqrt{\pi} a^2\gamma^{D+1}}\frac{\Gamma(D+1)}{2^{D+1}}\frac{\zeta_R(D+1)}{(D-2)!}\frac{\Gamma\left(\frac{D-1}{2}\right)}{\Gamma\left(\frac{D}{2}\right)}\\
=&(D-1)\times \frac{A(S^{D-1})}{d^{D+1}}\frac{D\Gamma\left(\frac{D+1}{2}\right)\zeta_R(D+1)}{2^{D+1}\pi^{\frac{D+1}{2}}},
\end{split}\end{equation}where $$A(S^{D-1})=\frac{2\pi^{\frac{D}{2}}}{\Gamma\left(\frac{D}{2}\right)}a^{D-1}$$ is the surface area of the sphere of radius $a$. For infinitely permeable boundary condition, the same derivation shows that the leading term of the Casimir force when $\gamma$ is small is also given by \eqref{eq3_26_7}. If we consider the Casimir force acting on the outer spherical shell, then the leading term is given by the negative of \eqref{eq3_26_7}, which signifies a force in the opposite direction. It is well-known that (see e.g. \cite{2} and \cite{37})
$$\frac{D\Gamma\left(\frac{D+1}{2}\right)\zeta_R(D+1)}{2^{D+1}\pi^{\frac{D+1}{2}}}\frac{1}{d^{D+1}}$$ is the magnitude of the attractive Casimir force per unit area acting between two parallel plates due to massless scalar field with Dirichlet boundary condition. Multiplying by the factor $(D-1)$ gives the magnitude of the Casimir force density due to electromagnetic field, where the factor $(D-1)$ accounts for the different polarizations   of photon in $D$ dimensional space. Therefore the analysis in this section shows that in the limit $\gamma=(b-a)/a$ is small, one finds that the leading term of the Casimir force per unit area between two concentric spherical shells is the Casimir force per unit area acting on a pair of parallel plates. This is a verification of the proximity force approximation \cite{39, 40}.

The lower order corrections to the proximity force approximation can be computed from \eqref{eq3_26_5} by expanding $\displaystyle \eta\left(\tfrac{bz}{a}\right)-\eta(z)$ and $p_{ni}(z;a,b)$ in terms of $\gamma$. The computation is quite involved and we leave it to the interested readers.

\section{Conclusion}\label{s9}This work considers electromagnetic field in $D$-dimensional spherically symmetric cavities subject to perfectly conducting or infinitely permeable boundary conditions.
We   derive  explicitly the eigenmodes of electromagnetic field in $D$-dimensional spherical coordinates. The field  modes are divided into TE modes and TM modes, and it is shown that TE modes can be decomposed into $(D-2)$ polarizations, whereas TM modes only have one polarization. In total, there are $(D-1)$ polarizations as one should expect. Although perfectly conducting and infinitely permeable boundary conditions are defined in terms of the two form $F_{\mu\nu}dx^{\mu}dx^{\nu}$ which represents the strength of the field, it is shown that in $D$-dimensional spherically symmetric cavities, the perfectly conducting   and infinitely permeable boundary conditions are equivalent respectively  to the relative and absolute boundary conditions on the potential one form $A=A_{\mu}dx^{\mu}$ defined by Gilkey \cite{18}.

The Casimir energies in the inside and outside of a spherical shell are computed using both cut-off regularization and zeta regularization. The sum of the inside and outside contributions is the Casimir energy of a single spherical shell system. It is observed that when the space dimension is $D=3$, the Casimir energy computed using either cut-off regularization or zeta regularization are  divergence free  and hence agree.  When $D\geq 4$,   divergences always present in cut-off regularization. However, using zeta regularization,   divergences are only present when $D$ is even, as is the case for massless scalar field. The heat kernel coefficients that appear in the divergences in the cut-off regularized Casimir energy are tabulated for $3\leq D\leq 9$. Casimir energies of the interior and exterior, and the Casimir energy of a single shell are also tabulated for $3\leq D\leq 9$. The results disagree with those tabulated in \cite{16,32,1} for the case of perfectly conducting boundary conditions since the TE modes are equated to modes of massless scalar field with Dirichlet boundary conditions in \cite{16,32,1}, which does not give rise to the correct number of polarizations. However, when $3\leq D\leq 7$, the values of the zeta regularized Casimir energies inside the spherical shell agrees with those computed in \cite{17}.

For the system of two spherical shells with radii $a$ and $b$, the total Casimir energy can be written as a sum of the single shell contribution from the inner shell, the single  shell contribution from the outer shell, and an interacting term. It is  shown that the interacting term always gives rise to a Casimir force that attracts the two shells to each other. When the separation between the shells $d=b-a$ is much smaller than the radii of the shells, the Casimir force   behaves like $(d/a)^{-D-1}$.  The leading term of the Casimir force per unit area is equal to the Casimir force per unit area acting between two parallel plates, as what one would expect from proximity force approximation.

\appendix
\section{Zeta Functions  }\label{a1} In this section, we compute the  zeta functions   that appear in the expressions of the Casimir energies. We follow the same approach as in \cite{30,31,15,17,32}.
\subsection{Zeta function for Casimir energy inside a spherical shell}
For Casimir energy inside a spherical shell, the corresponding zeta function is
\begin{equation}\label{eq3_3_5}
\begin{split}
\zeta^{\text{int}} (s)=\zeta_{\text{TE}}^{\text{int}} (s)+\zeta_{\text{TM}}^{\text{int}} (s)
=\sum_{l=1}^{\infty}h_D(l)\sum_{j=1}^{\infty}\left(\omega_{lj}^{\text{TE}}\right)^{-2s}+
 \sum_{l=1}^{\infty}b_D(l)\sum_{j=1}^{\infty}\left(\omega_{lj}^{\text{TM}}\right)^{-2s},
\end{split}
\end{equation}where $\omega_{lj}^{\text{TE}}$ and $\omega_{lj}^{\text{TM}}$ are the TE and TM modes (see Table \ref{t1} and Table \ref{t1_2}). Using residue theorem, we find that for perfectly conducting boundary condition,
\begin{equation}\label{eq3_24_1}
\zeta_{\text{TE}}^{\text{int}} (s)=\sum_{l=1}^{\infty}h_D(l)\zeta_D^{\text{int}, l+\frac{D-2}{2}}(s),\hspace{1cm}
\zeta_{\text{TM}}^{\text{int}} (s)=\sum_{l=1}^{\infty}b_D(l)\zeta_{R,\frac{D-2}{2}}^{\text{int}, l+\frac{D-2}{2}}(s),
\end{equation}and for infinitely permeable boundary condition,
\begin{equation}\label{eq3_24_2}
\zeta_{\text{TE}}^{\text{int}} (s)=\sum_{l=1}^{\infty}h_D(l)\zeta_{R,\frac{4-D}{2}}^{\text{int}, l+\frac{D-2}{2}}(s),\hspace{1cm}
\zeta_{\text{TM}}^{\text{int}} (s)=\sum_{l=1}^{\infty}b_D(l)\zeta_{D}^{\text{int}, l+\frac{D-2}{2}}(s).
\end{equation}Here
\begin{equation*}
\begin{split}
\zeta_D^{\text{int},\nu}(s)=&\lim_{m\rightarrow 0}\frac{1}{2\pi i}\oint_{\gamma}(z^2+m^2)^{-s} \frac{d}{dz}\log\Bigl\{z^{-\nu} J_{\nu}(az) \Bigr\}dz,\\
\zeta_{R,c}^{\text{int},\nu}(s)=&\lim_{m\rightarrow 0}\frac{1}{2\pi i}\oint_{\gamma}(z^2+m^2)^{-s} \frac{d}{dz}\log\Bigl\{z^{-\nu} \left[cJ_{\nu}(az)+azJ_{\nu}'(az)\right]
 \Bigr\}dz,
\end{split}
\end{equation*}where $\gamma$ is the contour consists of the line $x+i\varepsilon,   x=\infty \rightarrow 0$, followed by the semicircle $\displaystyle \varepsilon e^{i\theta}, \theta=\frac{\pi}{2}\rightarrow\frac{3\pi}{2}$, followed by the line $x-i\varepsilon,   x=0\rightarrow \infty$. The factor $z^{-\nu}$ is introduced to cancel the zeros of $J_{\nu}(z)$ at $z=0$.  The constant $m>0$ is introduced to simplify the manipulation of the integral. At the end,  $m$ is set to zero.  Using the fact that $\displaystyle J_{\nu}(iz)=e^{\frac{\pi \nu i}{2}}I_{\nu}(z), J_{\nu}(-iz)=e^{-\frac{\pi \nu i}{2}}I_{\nu}(z) $, where $I_{\nu}(z)$ is  the modified Bessel functions of the first kind   (see \cite{26}), one finds that
\begin{equation*}
\begin{split}
\zeta_D^{\text{int},\nu}(s)=&\lim_{m\rightarrow 0}\frac{\sin (\pi s)}{\pi}\int_m^{\infty}(z^2-m^2)^{-s} \frac{d}{dz}\log\Bigl\{ z^{-\nu} I_{\nu}(az) \Bigr\}dz
\\=&\lim_{m\rightarrow 0}\frac{\sin (\pi s)}{\pi}\int_{\frac{am}{\nu}}^{\infty}\left(\left(\frac{\nu z}{a}\right)^2-m^2\right)^{-s} \frac{d}{dz}\log\Bigl\{ z^{-\nu} I_{\nu}(\nu z) \Bigr\}dz,\\
\zeta_{R,c}^{\text{int}, \nu}(s)=&\lim_{m\rightarrow 0}\frac{\sin (\pi s)}{\pi }\int_m^{\infty}(z^2-m^2)^{-s}  \frac{d}{dz}\log\Bigl\{ z^{-\nu} \left[cI_{\nu}(az)+azI_{\nu}'(az)\right] \Bigr\}dz\\
=&\lim_{m\rightarrow 0}\frac{\sin (\pi s)}{\pi}\int_{\frac{am}{\nu}}^{\infty}\left(\left(\frac{\nu z}{a}\right)^2-m^2\right)^{-s}
 \frac{d}{dz}\log\Bigl\{ z^{-\nu} \left[cI_{\nu}(\nu z)+\nu zI_{\nu}'(\nu z)\right] \Bigr\}dz,
\end{split}
\end{equation*}for $\displaystyle \frac{1}{2}<\text{Re}\,s <1$.
To obtain the analytic continuation of $\zeta_D^{\text{int},\nu}(s)$ and $\zeta_{R,c}^{\text{int}, \nu}(s)$, one needs the Debye uniform asymptotic expansions of the Bessel function $I_{\nu}(\nu z)$ and $I_{\nu}'(\nu z)$  \cite{33, 34}: For fixed $z$, as $\nu\rightarrow \infty$,
\begin{equation}\label{eq3_4_1}\begin{split}
I_{\nu}(\nu z) \sim &\frac{1}{\sqrt{2\pi \nu}}\frac{e^{\nu\eta(z)}}{(1+z^2)^{1/4}} \sum_{k=0}^{\infty} \frac{u_k(t(z))}{\nu^k},\\
I_{\nu}'(\nu z)\sim & \frac{1}{\sqrt{2\pi \nu}}\frac{e^{\nu\eta(z)}(1+z^2)^{1/4}}{z}\sum_{k=0}^{\infty} \frac{v_k(t(z))}{\nu^k},
\end{split}\end{equation}where
\begin{align*}
\eta(z)=\sqrt{1+z^2}+\log\frac{z}{1+\sqrt{1+z^2}},\hspace{1cm}t(z)=\frac{1}{\sqrt{1+z^2}},
\end{align*}and $u_k(t)$ and $v_k(t)$ are defined recursively by
\begin{align*}
&u_0(t)=1, \hspace{0.5cm}u_{k}(t)=\frac{t^2(1-t^2)}{2}u_{k-1}'(t)+\frac{1}{8}\int_0^t(1-5\tau^2)u_{k-1}(\tau)d\tau,\\
&v_0(t)=1,\hspace{0.5cm} v_{k }(t)=u_k(t)-t^2(1-t^2)u_{k-1}'(t)-\frac{t(1-t^2)}{2}u_{k-1}(t).
\end{align*}
It can be shown that $u_k(t)$ and $v_k(t)$ are polynomials in $t$ of the form $\displaystyle \sum_{j=0}^k a_j t^{k+2j}$. For $cI_{\nu}(\nu z)+\nu zI_{\nu}'(\nu z)$,
\eqref{eq3_4_1} implies that
\begin{equation*}
cI_{\nu}(\nu z)+\nu zI_{\nu}'(\nu z)\sim  \frac{\sqrt{\nu}e^{\nu\eta(z)}(1+z^2)^{1/4}}{\sqrt{2\pi}}\left[1+\sum_{k=1}^{\infty} \frac{ct(z)u_{k-1}(t(z))+v_k(t(z))}{\nu^k}\right].
\end{equation*}Let
\begin{equation}\label{eq3_4_5}
\begin{split}
\log\left[1+\sum_{k=1}^{\infty} \frac{u_k(t )}{\nu^k}\right]=\sum_{n=1}^{\infty}\frac{D_n(t)}{\nu^n},\hspace{1cm}\log
\left[1+\sum_{k=1}^{\infty} \frac{ct u_{k-1}(t )+v_k(t )}{\nu^k}\right]=\sum_{n=1}^{\infty} \frac{M_{n,c}(t)}{\nu^n}.
\end{split}
\end{equation}It can be shown that $D_n(t)$ and $M_n(t)$ are polynomials of $t$ of the form:
\begin{align*}
D_n(t)=\sum_{j=0}^n d_{n,j}t^{n+2j}, \hspace{1cm}M_n(t)=\sum_{j=0}^n m_{n,j}(c)t^{n+2j}.
\end{align*}
From \eqref{eq3_4_1} and \eqref{eq3_4_5}, we have
\begin{equation}\label{eq3_4_2}
\begin{split}
&\frac{d}{dz}\log\Bigl\{z^{-\nu}I_{\nu}(\nu z)\Bigr\} \\=& -\frac{\nu}{z}+\frac{d}{dz} \log \left(e^{\nu\eta(z)}\right)-\frac{d}{dz} \log(1+z^2)^{\frac{1}{4}}+\frac{d}{dz}\sum_{i=1}^N
\frac{D_i(t(z))}{\nu^i} +\frac{d}{dz}\left(\log I_{\nu}(\nu z)-\log\frac{e^{\nu\eta(z)}}{\sqrt{2\pi \nu}(1+z^2)^{1/4}} -\sum_{i=1}^N \frac{D_i(t(z))}{\nu^i}\right)\\
=&\frac{\nu}{z}\left(\sqrt{1+z^2}-1\right)-\frac{1}{2}\frac{z}{1+z^2}-\sum_{i=1}^N \frac{zt(z)^3D_{i}'(t(z))}{\nu^i}+\frac{d}{dz}\left(\log I_{\nu}(\nu z)-\log\frac{e^{\nu\eta(z)}}{\sqrt{2\pi \nu}(1+z^2)^{1/4}} -\sum_{i=1}^N \frac{D_i(t(z))}{\nu^i}\right).
\end{split}
\end{equation}Similarly,
\begin{equation}\label{eq3_4_3}\begin{split}
 \frac{d}{dz}\log\Bigl\{z^{-\nu}\left[cI_{\nu}(\nu z)+\nu zI_{\nu}'(\nu z)\right]\Bigr\}
\sim & \frac{\nu}{z}\left(\sqrt{1+z^2}-1\right)+\frac{1}{2}\frac{z}{1+z^2}-\sum_{i=1}^N \frac{zt(z)^3M_{i,c}'(t(z))}{\nu^i}\\&+\frac{d}{dz}\left(\log \left[cI_{\nu}(\nu z)+\nu zI_{\nu}'(\nu z)\right]-\log\frac{\sqrt{\nu}e^{\nu\eta(z)}(1+z^2)^{1/4}}{\sqrt{2\pi}} -\sum_{i=1}^N \frac{M_{i,c}(t(z))}{\nu^i}\right).
\end{split}\end{equation}Here $N$ is any positive integer.  Using \eqref{eq3_4_2} and \eqref{eq3_4_3}, we can write the zeta functions $\zeta_D^{\text{int},\nu}(s)$ and
$\zeta_{R,c}^{\text{int},\nu}(s)$ as
\begin{equation*}
\begin{split}
 \zeta_D^{\text{int},\nu}(s)=\sum_{i=-1}^N A_{D,i}^{\text{int},\nu}(s)+B_{D,N}^{\text{int},\nu}(s),\hspace{1cm}
  \zeta_{R,c}^{\text{int},\nu}(s)=\sum_{i=-1}^N A_{R,c,i}^{\text{int},\nu}(s)+B_{R,c,N}^{\text{int},\nu}(s),
\end{split}
\end{equation*}where
\begin{equation*}
\begin{split}
A_{D,-1}^{\text{int},\nu}(s)=&A_{R,c,-1}^{\text{int},\nu}(s)=\lim_{m\rightarrow 0}\frac{\sin (\pi s)}{\pi}\int_{\frac{a m}{\nu}}^{\infty}\left(\left[\frac{\nu z}{a}\right]^2-m^2\right)^{-s}\left\{\frac{\nu}{z}\left(\sqrt{1+z^2}-1\right)\right\}dz
=\frac{a^{2s}\nu^{1-2s}}{4\sqrt{\pi}}\frac{\Gamma\left(s-\frac{1}{2}\right)}{\Gamma(s+1)},
\\
-A_{D,0}^{\text{int},\nu}(s)=&A_{R,c,0}^{\text{int},\nu}(s)=\lim_{m\rightarrow 0}\frac{\sin (\pi s)}{\pi}\int_{\frac{a m}{\nu}}^{\infty}\left(\left[\frac{\nu z}{a}\right]^2-m^2\right)^{-s}\left\{\frac{1}{2}\frac{z}{1+z^2}\right\}dz=\frac{a^{2s}\nu^{-2s}}{4};
\end{split}
\end{equation*}  for $i\geq 1$,
\begin{equation*}\begin{split}
A_{D,i}^{\text{int},\nu}(s) =&-\lim_{m\rightarrow 0}\frac{\sin (\pi s)}{\pi}\int_{\frac{a m}{\nu}}^{\infty}\left(\left[\frac{\nu z}{a}\right]^2-m^2\right)^{-s}\left\{\frac{zt(z)^3D_{i}'(t(z))}{\nu^i}\right\}dz
=-a^{2s}\nu^{-2s-i}\sum_{j=0}^{i}d_{i,j}\frac{\Gamma\left(s+\frac{i+2j}{2}\right)}{\Gamma(s)\Gamma\left(\frac{i+2j}{2}\right)},\\
A_{R,c,i}^{\text{int},\nu}(s) =&-\lim_{m\rightarrow 0}\frac{\sin (\pi s)}{\pi}\int_{\frac{a m}{\nu}}^{\infty}\left(\left[\frac{\nu z}{a}\right]^2-m^2\right)^{-s}\left\{\frac{zt(z)^3M_{i,c}'(t(z))}{\nu^i}\right\}dz
=-a^{2s}\nu^{-2s-i}\sum_{j=0}^{i}m_{i,j}(c)\frac{\Gamma\left(s+\frac{i+2j}{2}\right)}{\Gamma(s)\Gamma\left(\frac{i+2j}{2}\right)};\end{split}
\end{equation*}
and
\begin{align*}
B_{D,N}^{\text{int},\nu}(s)=&\lim_{m\rightarrow 0}\frac{\sin (\pi s)}{\pi}\int_{\frac{a m}{\nu}}^{\infty}\left(\left[\frac{\nu z}{a}\right]^2-m^2\right)^{-s}
\frac{d}{dz}\left(\log I_{\nu}(\nu z)-\log\frac{e^{\nu\eta(z)}}{\sqrt{2\pi \nu}(1+z^2)^{1/4}} -\sum_{i=1}^N \frac{D_i(t(z))}{\nu^i}\right)dz\\
B_{R,c,N}^{\text{int},\nu}(s)=&\lim_{m\rightarrow 0}\frac{\sin (\pi s)}{\pi}\int_{\frac{a m}{\nu}}^{\infty}\left(\left[\frac{\nu z}{a}\right]^2-m^2\right)^{-s}
\\&\hspace{4cm}\times\frac{d}{dz}\left(\log \left[cI_{\nu}(\nu z)+\nu zI_{\nu}'(\nu z)\right]-\log\frac{\sqrt{\nu}e^{\nu\eta(z)}(1+z^2)^{1/4}}{\sqrt{2\pi}} -\sum_{i=1}^N \frac{M_{i,c}(t(z))}{\nu^i}\right)dz.
\end{align*}
Now notice that $b_D(l)$ and $h_D(l)$ can be expanded as
\begin{equation*}\begin{split}
h_{D}(l)=&\frac{l(l+D-2)(2l+D-2)(l+D-4)!}{(D-3)!(l+1)!} =\sum_{j=0}^{ D-2} x_{D;j}\left(l+\frac{D-2}{2}\right)^j,\\
 b_D(l) = & \frac{(2l+D-2)(l+D-3)!}{(D-2)! l!}=\sum_{j=1}^{ D-2} y_{D;j}\left(l+\frac{D-2}{2}\right)^j,\end{split}
\end{equation*}where $x_{D;j}=0$ and $y_{D;j}=0$ if $j$ and $D$ have opposite parity, and $x_{D;0}\neq 0$ if and only if $D=4$. Let
\begin{equation}\label{eq3_24_4}
\zeta_H(s;\chi)=\sum_{k=0}^{\infty}(k+\chi)^{-s}
\end{equation}be the Hurwitz zeta function. From \eqref{eq3_24_1} and \eqref{eq3_24_2}, we find that
in the case of perfectly conducting boundary condition, the zeta function \eqref{eq3_3_5} can be written as
\begin{equation}\label{eq3_8_1}
\begin{split}
\zeta^{\text{int}}(s)=&\sum_{j=0}^{D-2} x_{D;j} \sum_{l=1}^{\infty}\left(l+\tfrac{D-2}{2}\right)^j \sum_{i=-1}^N A_{D,i}^{\text{int},l+\tfrac{D-2}{2}}(s)
+\sum_{l=1}^{\infty}h_l(D) B_{D,N}^{\text{int},l+\tfrac{D-2}{2}}(s)\\&+ \sum_{j=1}^{D-2} y_{D;j} \sum_{l=1}^{\infty}\left(l+\tfrac{D-2}{2}\right)^j \sum_{i=-1}^N A_{R,\tfrac{D-2}{2},i}^{\text{int},l+\tfrac{D-2}{2}}(s)
+\sum_{l=1}^{\infty}b_l(D) B_{R,\frac{D-2}{3},N}^{\text{int},l+\tfrac{D-2}{2}}(s)\end{split}\end{equation}
\begin{equation*}\begin{split}
=&  \frac{a^{2s}}{4\sqrt{\pi}}\sum_{j=0}^{D-2}\left(x_{D;j}+y_{D;j}\right)\frac{\Gamma\left(s-\frac{1}{2}\right)}{\Gamma(s+1)}\zeta_H\left(2s-j-1;\tfrac{D}{2}\right)
-\frac{a^{2s}}{4}\sum_{j=0}^{D-2}\left(x_{D;j}-y_{D;j}\right)\zeta_H\left( 2s-j;\tfrac{D}{2}\right)\\&-a^{2s}\sum_{j=0}^{D-2}x_{D;j}\sum_{i=1}^N\zeta_H\left(2s+i-j;\tfrac{D}{2}\right)\sum_{k=0}^id_{i,k}\frac{\Gamma\left(s+\frac{i+2k}{2}\right)}
{\Gamma(s)\Gamma\left(\frac{i+2k}{2}\right)} \\&-a^{2s}\sum_{j=1}^{D-2}y_{D;j}\sum_{i=1}^N\zeta_H\left(2s+i-j;\tfrac{D}{2}\right)\sum_{k=0}^im_{i,k}\left(\tfrac{D-2}{2}\right)\frac{\Gamma\left(s+\frac{i+2k}{2}\right)}
{\Gamma(s)\Gamma\left(\frac{i+2k}{2}\right)}  +B^{\text{int}}_N(s),
\end{split}
\end{equation*}where
\begin{equation*}\begin{split}
B^{\text{int}}_N(s)=&a^{2s}\frac{\sin (\pi s)}{\pi} \sum_{l=1}^{\infty}h_D(l)\nu_D(l)^{-2s} \int_{0}^{\infty}z^{-2s}
\frac{d}{dz}\Biggl(\log I_{\nu_D(l)}\left(\nu_D(l)z\right)  -\log\frac{e^{\nu_D(l)\eta(z)}}{\sqrt{2\pi \nu_D(l)}(1+z^2)^{1/4}} -\sum_{i=1}^N \frac{D_i(t(z))}{\nu_D(l)^i}\Biggr)dz\\
&+a^{2s}\frac{\sin (\pi s)}{\pi} \sum_{l=1}^{\infty}b_D(l)\nu_D(l)^{ -2s} \int_{0}^{\infty}z^{-2s}
\frac{d}{dz}\Biggl(\log \left[\tfrac{D-2}{2}I_{\nu_D(l)}\left(\nu_D(l) z\right)+\nu_D(l) zI_{\nu_D(l)}'\left(\nu_D(l) z\right)\right]\\&\hspace{3cm}-\log\frac{\sqrt{ \nu_D(l)}e^{\nu_D(l)\eta(z)}(1+z^2)^{1/4}}{\sqrt{2\pi}} -\sum_{i=1}^N \frac{M_{i,\frac{D-2}{2}}(t(z))}{\nu_D(l)^i} \Biggr)dz.
\end{split}\end{equation*}Here $\displaystyle \nu_D(l)=l+\tfrac{D-2}{2}$.  The result for infinitely permeable boundary condition can be obtained in a similar way.
By taking $N\geq D$, one can guarantee that $B^{\text{int}}_N(s)$ does not contain any pole  on the half-plane $\text{Re}\;s>-1$.

\subsection{Zeta function for Casimir energy outside a spherical shell}
For Casimir energy outside a spherical shell, consider the zeta function:
\begin{equation}\label{eq3_15_1}
\begin{split}
\zeta^{\text{ext}} (s)=\zeta_{\text{TE}}^{\text{ext}} (s)+\zeta_{\text{TM}}^{\text{ext}} (s),
\end{split}
\end{equation}where for perfectly conducting boundary condition,
\begin{equation*}
\zeta_{\text{TE}}^{\text{ext}} (s)=\sum_{l=1}^{\infty}h_D(l)\zeta_D^{\text{ext}, l+\frac{D-2}{2}}(s),\hspace{1cm}
\zeta_{\text{TM}}^{\text{ext}} (s)=\sum_{l=1}^{\infty}b_D(l)\zeta_{R,\frac{D-2}{2}}^{\text{ext}, l+\frac{D-2}{2}}(s),
\end{equation*}and for infinitely permeable boundary condition,
\begin{equation*}
\zeta_{\text{TE}}^{\text{ext}} (s)=\sum_{l=1}^{\infty}h_D(l)\zeta_{R,\frac{4-D}{2}}^{\text{ext}, l+\frac{D-2}{2}}(s),\hspace{1cm}
\zeta_{\text{TM}}^{\text{ext}} (s)=\sum_{l=1}^{\infty}b_D(l)\zeta_{D}^{\text{ext}, l+\frac{D-2}{2}}(s).
\end{equation*}Here
\begin{equation*}
\begin{split}
\zeta_D^{\text{ext},\nu}(s)=&\lim_{m\rightarrow 0}\frac{\sin (\pi s)}{\pi}\int_{\frac{a m}{\nu}}^{\infty}\left(\left[\frac{\nu z}{a}\right]^2-m^2\right)^{-s} \frac{d}{dz}\log\Bigl\{ z^{-\nu} K_{\nu}(\nu z) \Bigr\}dz,\\
\zeta_{R,c}^{\text{ext}, \nu}(s)=&\lim_{m\rightarrow 0}\frac{\sin (\pi s)}{\pi }\int_{\frac{a m}{\nu}}^{\infty}\left(\left[\frac{\nu z}{a}\right]^2-m^2\right)^{-s} \frac{d}{dz}\log\Bigl\{ z^{-\nu} \left[-cK_{\nu}(\nu z)-\nu zK_{\nu}'(\nu z)\right] \Bigr\}dz,
\end{split}
\end{equation*}for $\displaystyle \frac{1}{2}<\text{Re}\,s <1$. For the Bessel function $K_{\nu}(\nu z)$ and its derivative, the Debye uniform asymptotic expansions read as
\cite{33,34}: For fixed $z$, as $\nu\rightarrow \infty$,
\begin{equation}\label{eq3_15_2}\begin{split}
K_{\nu}(\nu z) \sim &\sqrt{\frac{\pi}{2 \nu}}\frac{e^{-\nu\eta(z)}}{(1+z^2)^{1/4}} \sum_{k=0}^{\infty} (-1)^k\frac{u_k(t(z))}{\nu^k},\\
K_{\nu}'(\nu z)\sim & -\sqrt{\frac{\pi}{2 \nu}}\frac{e^{-\nu\eta(z)}(1+z^2)^{1/4}}{z}\sum_{k=0}^{\infty}(-1)^k \frac{v_k(t(z))}{\nu^k}.
\end{split}\end{equation}The function $u_k(t), v_k(t), \eta(z), t(z)$ are defined in the previous section. Eq. \eqref{eq3_15_2} implies that
\begin{equation*}
cK_{\nu}(\nu z)+\nu zK_{\nu}'(\nu z)\sim - \sqrt{\frac{\pi\nu}{2}}  e^{-\nu\eta(z)}(1+z^2)^{1/4}\left[1+\sum_{k=1}^{\infty}(-1)^k \frac{ct(z)u_{k-1}(t(z))+v_k(t(z))}{\nu^k}\right].
\end{equation*}Since
\begin{equation}\label{eq3_15_3}
\begin{split}
\log\left[1+\sum_{k=1}^{\infty} (-1)^k\frac{u_k(t )}{\nu^k}\right]=\sum_{n=1}^{\infty}(-1)^n\frac{D_n(t)}{\nu^n},\hspace{1cm}\log
\left[1+\sum_{k=1}^{\infty}(-1)^k \frac{ct u_{k-1}(t )+v_k(t )}{\nu^k}\right]=\sum_{n=1}^{\infty} (-1)^n\frac{M_{n,c}(t)}{\nu^n},
\end{split}
\end{equation}
the same approach as in the previous section shows that
\begin{equation*}
\begin{split}
 \zeta_D^{\text{ext},\nu}(s)=\sum_{i=-1}^N A_{D,i}^{\text{ext},\nu}(s)+B_{D,N}^{\text{ext},\nu}(s),\hspace{1cm}
  \zeta_{R,c}^{\text{ext},\nu}(s)=\sum_{i=-1}^N A_{R,c,i}^{\text{ext},\nu}(s)+B_{R,c,N}^{\text{ext},\nu}(s),
\end{split}
\end{equation*}where
\begin{equation*}
A_{D,i}^{\text{ext},\nu}(s)=(-1)^iA_{D,i}^{\text{int},\nu}(s),\hspace{1cm}A_{R,c,i}^{\text{ext},\nu}(s)=(-1)^iA_{R,c,i}^{\text{int},\nu}(s),
\end{equation*}and
\begin{align*}
B_{D,N}^{\text{ext},\nu}(s)=&\lim_{m\rightarrow 0}\frac{\sin (\pi s)}{\pi}\int_{\frac{a m}{\nu}}^{\infty}\left(\left[\frac{\nu z}{a}\right]^2-m^2\right)^{-s}
\frac{d}{dz}\left(\log K_{\nu}(\nu z)-\log\frac{\sqrt{\pi}e^{-\nu\eta(z)}}{\sqrt{2  \nu}(1+z^2)^{1/4}} -\sum_{i=1}^N (-1)^i\frac{D_i(t(z))}{\nu^i}\right)dz\\
B_{R,c,N}^{\text{int},\nu}(s)=&\lim_{m\rightarrow 0}\frac{\sin (\pi s)}{\pi}\int_{\frac{a m}{\nu}}^{\infty}\left(\left[\frac{\nu z}{a}\right]^2-m^2\right)^{-s}
\\&\hspace{2cm}\times\frac{d}{dz}\left(\log \left[-cK_{\nu}(\nu z)-\nu zK_{\nu}'(\nu z)\right]-\log\frac{\sqrt{\pi \nu}e^{-\nu\eta(z)}(1+z^2)^{1/4}}{\sqrt{2 }} -\sum_{i=1}^N (-1)^i\frac{M_{i,c}(t(z))}{\nu^i}\right)dz.
\end{align*}
From these, we find that for perfectly conducting boundary condition,
\begin{equation}\label{eq3_24_7}
\begin{split}
\zeta^{\text{ext}}(s)
=&  -\frac{a^{2s}}{4\sqrt{\pi}}\sum_{j=0}^{D-2}\left(x_{D;j}+y_{D;j}\right)\frac{\Gamma\left(s-\frac{1}{2}\right)}{\Gamma(s+1)}\zeta_H\left(2s-j-1;\tfrac{D}{2}\right)
-\frac{a^{2s}}{4}\sum_{j=0}^{D-2}\left(x_{D;j}-y_{D;j}\right)\zeta_H\left( 2s-j;\tfrac{D}{2}\right)\\&-a^{2s}\sum_{j=0}^{D-2}x_{D;j}\sum_{i=1}^N(-1)^i\zeta_H\left(2s+i-j;\tfrac{D}{2}\right)\sum_{k=0}^id_{i,k}\frac{\Gamma\left(s+\frac{i+2k}{2}\right)}
{\Gamma(s)\Gamma\left(\frac{i+2k}{2}\right)} \\&-a^{2s}\sum_{j=1}^{D-2}y_{D;j}\sum_{i=1}^N(-1)^i\zeta_H\left(2s+i-j;\tfrac{D}{2}\right)\sum_{k=0}^im_{i,k}\left(\tfrac{D-2}{2}\right)\frac{\Gamma\left(s+\frac{i+2k}{2}\right)}
{\Gamma(s)\Gamma\left(\frac{i+2k}{2}\right)}  +B^{\text{ext}}_N(s),
\end{split}
\end{equation}where
\begin{equation*}\begin{split}
&B^{\text{ext}}_N(s)\\=&a^{2s}\frac{\sin (\pi s)}{\pi} \sum_{l=1}^{\infty}h_D(l)^{ -2s} \int_{0}^{\infty}z^{-2s}
\frac{d}{dz}\Biggl(\log K_{\nu_D(l)}\left(\nu_D(l)z\right)  -\log\frac{\sqrt{\pi}e^{-\nu_D(l)\eta(z)}}{\sqrt{2  \nu_D(l)}(1+z^2)^{1/4}}  -\sum_{i=1}^N (-1)^i \frac{D_i(t(z))}{\nu_D(l)^i}\Biggr)dz \\&
 +a^{2s}\frac{\sin (\pi s)}{\pi} \sum_{l=1}^{\infty}b_D(l)^{ -2s} \int_{0}^{\infty}z^{-2s}
 \frac{d}{dz}\Biggl(\log \left[-\tfrac{D-2}{2}K_{\nu_D(l)}\left(\nu_D(l) z\right)-\nu_D(l) zK_{\nu_D(l)}'\left(\nu_D(l) z\right)\right]\\&\hspace{3cm}-\log\frac{\sqrt{ \pi \nu_D(l)}e^{-\nu_D(l)\eta(z)}(1+z^2)^{1/4}}{\sqrt{2 }} -\sum_{i=1}^N (-1)^i\frac{M_{i,\frac{D-2}{2}}(t(z))}{\nu_D(l)^i} \Biggr)dz.
\end{split}\end{equation*}Here $\displaystyle \nu_D(l)=l+\tfrac{D-2}{2}$ as before. It is easy to write down the corresponding expressions for infinitely permeable boundary conditions.
\subsection{Zeta function for the interacting part of the Casimir energy in an annular region}
As defined in Section \ref{s4}, the zeta function that corresponds to the interacting term of the Casimir energy inside an annular region bounded by spheres of radius $a$ and $b$ is given by
\begin{equation}\label{eq3_15_1}
\begin{split}
\zeta^{\text{inter}} (s)=\zeta_{\text{TE}}^{\text{inter}} (s)+\zeta_{\text{TM}}^{\text{inter}} (s),
\end{split}
\end{equation}where for perfectly conducting boundary conditions,
\begin{equation*}
\zeta_{\text{TE}}^{\text{inter}} (s)=\sum_{l=1}^{\infty}h_D(l)\zeta_D^{\text{inter}, l+\frac{D-2}{2}}(s),\hspace{1cm}
\zeta_{\text{TM}}^{\text{inter}} (s)=\sum_{l=1}^{\infty}b_D(l)\zeta_{R,\frac{D-2}{2}}^{\text{inter}, l+\frac{D-2}{2}}(s),
\end{equation*}and for infinitely permeable boundary conditions,
\begin{equation*}
\zeta_{\text{TE}}^{\text{inter}} (s)=\sum_{l=1}^{\infty}h_D(l)\zeta_{R,\frac{4-D}{2}}^{\text{inter}, l+\frac{D-2}{2}}(s),\hspace{1cm}
\zeta_{\text{TM}}^{\text{inter}} (s)=\sum_{l=1}^{\infty}b_D(l)\zeta_{D}^{\text{inter}, l+\frac{D-2}{2}}(s).
\end{equation*}Here
\begin{equation}\label{eq3_26_1}\begin{split}
\zeta_D^{\text{inter}, \nu}(s)=&\lim_{m\rightarrow 0}\frac{\sin (\pi s)}{\pi }\int_m^{\infty}(z^2-m^2)^{-s} \frac{d}{dz}\log\left\{ 1-\frac{I_{\nu}(az)K_{\nu}(bz)}{I_{\nu}(bz)K_{\nu}(az)}\right\}dz,\\
\zeta_{R,c}^{\text{inter}, \nu}(s)=&\lim_{m\rightarrow 0}\frac{\sin (\pi s)}{\pi }\int_m^{\infty}(z^2-m^2)^{-s} \frac{d}{dz}\log\left\{  1-\frac{\left[cI_{\nu}(az)+azI_{\nu}'(az)\right]\left[cK_{\nu}(bz)+bzK_{\nu}'(bz)\right]}{\left[cI_{\nu}(bz)+bzI_{\nu}'(bz)\right]
\left[cK_{\nu}(az)+azK_{\nu}'(az)\right]} \right\}dz.
\end{split}\end{equation}The Debye uniform asymptotic expansions \eqref{eq3_4_1} and \eqref{eq3_15_2} for $I_{\nu}(\nu z)$ and $K_{\nu}(\nu z)$ show  that as $\nu\rightarrow \infty$,
\begin{equation}\label{eq3_25_1}
\begin{split}
\frac{I_{\nu}(\nu az)K_{\nu}(\nu bz)}{I_{\nu}(\nu bz)K_{\nu}(\nu az)}\sim \exp\left(-2\nu(\eta(bz)-\eta(az))\right)=\left(\frac{a(1+\sqrt{1+b^2z^2})}{b(1+\sqrt{1+a^2z^2})}\right)^{2\nu} \exp\left(-2\nu\left(\sqrt{1+b^2z^2}-\sqrt{1+a^2z^2}\right)\right).
\end{split}
\end{equation}The function $$\frac{a(1+\sqrt{1+b^2z^2})}{b(1+\sqrt{1+a^2z^2})}$$ is $<1$ for all $z$. Due to the power $2\nu$ on this function, we find that by setting $m=0$ directly, the first function  on the right hand side of \eqref{eq3_26_1} is analytic for all $\displaystyle \text{Re}\,s<\frac{1}{2}$. It can be analytically continued to the whole plane using standard tricks. The possible poles of the integral appear  at $s=1,2,\ldots$ are canceled by the zeros of $\sin(\pi s)$ at these points.   This shows that $\zeta_D^{\text{inter}, \nu}(s)$ is an analytic function for all $s$. Moreover, it is obvious  that as $b\rightarrow \infty$, the expression on the left hand side of \eqref{eq3_25_1} vanishes. Therefore,
\begin{align*}
\lim_{b\rightarrow \infty} \zeta_D^{\text{inter}, \nu}(s)=0.
\end{align*}Similar argument shows that $\zeta_{R,c}^{\text{inter}, \nu}(s)$ is an analytic function for all $s$, and
\begin{align*}
\lim_{b\rightarrow \infty} \zeta_{R,c}^{\text{inter}, \nu}(s)=0.
\end{align*}For the value of the zeta function $\zeta^{\text{inter}} (s)$ at $\displaystyle s=-\frac{1}{2}$, one can immediately set $\displaystyle s=-\frac{1}{2}$. Upon integration by parts, one finds that for perfectly conducting boundary conditions,
\begin{equation}\label{eq3_25_2}
\begin{split}
\zeta^{\text{inter}}\left(-\frac{1}{2}\right)=&\frac{1}{\pi}\sum_{l=1}^{\infty}h_D(l)\int_0^{\infty}\log\left\{ 1-\frac{I_{\nu(l)}(az)K_{\nu(l)}(bz)}{I_{\nu(l)}(bz)K_{\nu(l)}(az)}\right\}dz
\\&+\frac{1}{\pi}\sum_{l=1}^{\infty}b_D(l)\int_0^{\infty}\log\left\{  1-\frac{\left[\tfrac{D-2}{2}I_{\nu(l)}(az)+azI_{\nu(l)}'(az)\right]\left[\tfrac{D-2}{2}K_{\nu(l)}(bz)+bzK_{\nu(l)}'(bz)\right]}{\left[I_{\nu(l)}(bz)+bzI_{\nu(l)}'(bz)\right]
\left[\tfrac{D-2}{2}K_{\nu(l)}(az)+azK_{\nu(l)}'(az)\right]} \right\}dz.
\end{split}
\end{equation} It is straightforward to write down the corresponding expression for infinitely permeable boundary conditions.

\end{document}